\documentclass[journal]{IEEEtran}
\usepackage{verbatim}  
\usepackage{mathrsfs}  
\usepackage{bm}        
\usepackage{chngpage}  
\usepackage{array}
\usepackage{url}       
\usepackage{booktabs}  
\usepackage{multirow}
\usepackage{textcomp}  
\usepackage[nodisplayskipstretch]{setspace}
\usepackage{cite}
\usepackage{amsmath}
\usepackage{cleveref}  
\usepackage{amssymb}   
\usepackage{makecell}
\usepackage{graphicx}
\usepackage{xcolor}
\usepackage{epstopdf}
\usepackage{etoolbox}
\usepackage{ulem}
\usepackage{algorithm}
\usepackage{algpseudocode}
\newcommand{\subparagraph}{}  
\usepackage{titlesec}
\usepackage{threeparttable}

\makegapedcells
\setcellgapes{0pt}

\ifCLASSOPTIONcompsoc
	\usepackage[caption=false,font=normalsize,labelfont=sf,textfont=sf,subrefformat=parens,labelformat=parens]{subfig}
\else
	\usepackage[caption=false,font=footnotesize,subrefformat=parens,labelformat=parens]{subfig}
\fi

\newcommand{\jiang}{\fontsize{9.5pt}{11.0pt}\selectfont}

\makeatletter
\patchcmd{\@maketitle}
{\addvspace{0.5\baselineskip}\egroup}
{\addvspace{-2.0\baselineskip}\egroup}
{}
{}
\makeatother

\makeatletter
\def\subsubsection{%
  \@startsection
    {subsubsection}                 
    {3}                             
    {\parindent}                    
    {0ex plus 0.1ex minus 0.1ex}    
    {0ex}                           
    {\normalfont\normalsize\itshape}
}%
\makeatother

\titlespacing{\section}{0pt}{0.10\baselineskip}{0.02\baselineskip}
\titlespacing{\subsection}{0pt}{0.02\baselineskip}{0.02\baselineskip}

\begin{document}
\title{Convex Relaxation of Combined Heat\\and Power Dispatch}

\author{
Yibao~Jiang,~\IEEEmembership{Student Member,~IEEE,}
Can~Wan,~\IEEEmembership{Member,~IEEE,}
Audun~Botterud,~\IEEEmembership{Member,~IEEE,}
Yonghua~Song,~\IEEEmembership{Fellow,~IEEE,} 
and Mohammad Shahidehpour,~\IEEEmembership{Fellow,~IEEE} 
\thanks{This work was partially supported by National Natural Science Foundation 
of China (Grant No. 51761135015). \textit{(Corresponding author: Can Wan.)}} 
\thanks{Y. Jiang and C. Wan are with the College of Electrical Engineering, 
Zhejiang University, Hangzhou 310027, China 
(e-mail: jiangyb@zju.edu.cn, canwan@zju.edu.cn).} 
\thanks{A. Botterud is with the Laboratory for Information and Decision Systems, 
Massachusetts Institute of Technology, Cambridge, MA 02139 USA (e-mail: audunb@mit.edu).} 
\thanks{Y. Song is with State Key Laboratory of Internet of Things for Smart City, 
University of Macau, Taipa, Macau SAR, China, 
and also with the College of Electrical Engineering, Zhejiang University, Hangzhou 310027, 
China (e-mail: yhsong@um.edu.mo).} 
\thanks{M. Shahidehpour is with the Electrical and Computer Engineering Department, 
Illinois Institute of Technology, Chicago, IL 60616, USA (e-mail:ms@iit.edu).}} 


\maketitle

\normalem

\begin{abstract}
Combined heat and power dispatch promotes interactions and synergies 
between electric power systems and district heating systems.
However, nonlinear and nonconvex heating flow imposes significant challenges 
on finding qualified solutions efficiently. 
Most existing methods rely on constant flow assumptions 
to derive a linear heating flow model,
sacrificing optimality for computational simplicity. 
This paper proposes a novel convex combined heat and power dispatch model 
based on model simplification and constraint relaxation,
which improves solution quality and avoids assumptions on operating regimes of 
district heating systems.
To alleviate mathematical complexity introduced by the commonly used node method,
a simplified thermal dynamic model is proposed to capture temperature changes
in networked pipelines.
Conic and polyhedral relaxations are then applied to 
convexify the original problems with bilinear and quadratic equality constraints. 
Furthermore, an adaptive solution algorithm is proposed to successively reduce 
relaxation gaps based on dynamic bivariate partitioning, improving solution optimality
with desirable computational efficiency. The proposed method
is verified on a 33-bus electric power system integrated with
a 30-node district heating system and compared to nonlinear programming solvers 
and constant-flow-based solutions.
\end{abstract}

\vspace{-1.4mm}
\begin{IEEEkeywords}
Convex relaxation, combined heat and power dispatch, district heating flow,
integer programming.
\end{IEEEkeywords}

\jiang

\section{Introduction}

\IEEEPARstart{I}{ncreasing} deployment of combined heat and power plants (CHP) 
and heat pumps 
intensifies energy interactions between electric power systems (EPS) and 
district heating systems (DHS) \cite{dall2017unlocking,jiang2020hybrid}.
Combined heat and power dispatch (CHPD) is utilized for coordinated scheduling of 
interdependent electricity and heat systems, 
which promotes a minimization of operational costs 
and enhances renewable energy utilization \cite{rong2018dynamic,jiang2020exploiting}. 

In many parts of the world, heat is generated and supplied in a centralized way 
through networked pipelines with higher efficiency and less pollution,
such as in North China and Europe \cite{lund2010role}.
Thus, the process of water distribution and heat delivery needs to be modeled 
in CHPD. A combined analysis of electric power and heating flow is studied
in \cite{liu2016combined} based on 
static hydraulic-thermal conditions of DHS \cite{pirouti2013modelling}.
Network-constrained heating flow is incorporated into the simulation model, 
while time delays and temperature dynamics of heat distribution 
are not fully considered \cite{liu2016combined}.
Thermal dynamics in DHS are investigated in \cite{li2015combined} 
based on the so-called node method \cite{palsson1999equivalent}, 
which captures the correlation of nodal temperature at different time steps.
Thermal storage and network flexibility of DHS in CHPD are further explored in 
\cite{liu2019optimal,jiang2019combined,chen2014increasing,dai2018general} 
for cost reduction and integration of renewable power. 

One basic challenge so far is that nonlinear and nonconvex network flow 
renders CHPD problems hard to solve precisely and efficiently. 
Mathematically speaking, the CHPD problem considering electric power flow and
district heating flow is of nonconvex nonlinear programming (NLP) 
or mixed-integer nonlinear programming type, for which it is quite
difficult to obtain optimal solutions within limited time \cite{bertsekas1997nonlinear}.
Available approaches with respect to CHPD can be briefly categorized into two groups:
heuristic iterative algorithms \cite{wang2006stochastic,li2015combined}
or model assumptions \cite{liu2019optimal}.

In general, heuristic algorithms solve the CHPD problem by either metaheuristics
or customized iterations. Reference \cite{nazari2018comprehensive}
provides a review of metaheuristics implemented in 
optimal scheduling of power and heat generation, including
particle swarm optimization \cite{wang2006stochastic},
gravitational search algorithm, and so on.
However, theses metaheuristics do not include complex heating flow
as model constraints. 
On the other hand, some solution algorithms 
use iterative strategies by fixing time delays of heat delivery \cite{li2015combined}
or mass flow rates \cite{cao2018decentralized} in each iteration. 
These iterative strategies can obtain desired solutions by tuning 
relevant parameters to appropriate values. 
However, they tend to be specified for certain scenarios and rely on extensive 
experience in designing a viable strategy. Besides,
some existing algorithms \cite{li2015combined} rely on general NLP solvers 
(such as IPOPT \cite{wachter2006implementation}) 
to produce a candidate solution if both mass flow rates and 
nodal temperatures in the DHS are set as decision variables. 
NLP solvers based on interior point methods only aim to find local solutions 
for nonconvex problems, and may have unsatisfactory performance. 
Another commonly used method for CHPD employs model assumptions,
where DHS is simplified to the constant-flow mode \cite{liu2019optimal,wu2018combined}.
As a consequence, the computational complexity is significantly 
reduced since the heating flow model is linear under the constant-flow assumption.
However, this approach obviously sacrifices optimality for computing simplicity,
since many candidate solutions are potentially eliminated from the solution space.

Reduction in operational costs 
may save considerable amount of money depending on the system size. 
Therefore, it is critical to improve 
the solution quality and computational performance for CHPD. 
To this end, a convex CHPD model is proposed based on model simplification and 
constraint relaxation in this paper. 
The proposed convex model is designed for general CHPD problems 
without any assumptions on operating regimes. 
Convexity guarantees that a globally optimal solution with respect to 
the relaxed CHPD model can be found efficiently.
In particular, a simplified thermal dynamic (STD) model is developed to depict 
temperature changes in the DHS,
avoiding introducing numerous integer variables
that is required to setup temperature correlation for the node method.
Conic and polyhedral relaxations are then utilized 
to relax nonconvex quadratic equality constraints and  bilinear constraints.
The convex CHPD model is not equivalent to the original problem,
since relaxation enlarges the original nonconvex feasible set until it is convex.
It is thus critical to derive a tight relaxation to improve the solution quality.
Here, an adaptive solution algorithm is established to enhance the relaxation quality 
based on dynamic bivariate partitioning and piecewise McCormick envelopes \cite{bergamini2005logic}. 
Variable domains pertaining to bilinear terms are sequentially divided into a given
number of partitions to reduce the size of relaxed feasible regions. 
Instead of uniformly partitioning domains of all variables,
the adaptive strategy identifies bilinear constraints that are most significantly violated
and iteratively add incremental partitions
to avoid introducing too many binary variables.
Thus, the relaxation quality can be strengthened successively with satisfactory 
computational efficiency.
Numerical experiments on a test system composed of a 33-bus EPS and a 30-node DHS 
validate the effectiveness of the proposed method in terms of 
solution quality and computational feasibility.

To the authors' best knowledge, this is the first paper investigating 
convex optimization and solution algorithms for general CHPD problems.
The main contributions are summarized as follows:
1) A novel convex CHPD model is proposed without assumptions on constant mass flow rates.
The model jointly optimizes operational strategies for EPS and DHS with 
satisfactory computational efficiency and improved solution quality;
2) A simplified thermal dynamic model is developed to characterize temperature 
changes in heating networks, mitigating the mathematical complexity of 
the node method.
3) Both conic and polyhedral relaxations are implemented in the convex CHPD model
with thorough modeling of hydraulic and thermal conditions in the DHS.
4) A novel adaptive solution algorithm is devised 
based on dynamic bivariate partitioning and piecewise polyhedral relaxation
to reduce relaxation gaps while preserving computational efficiency.

\section{Mathematical Model of CHPD}
\label{sec:CHPDModel}

\subsection{Electric Power System}
Considering the limited size of DHS, electric power
distribution systems are modeled here with different types of distributed 
generators, including gas-fired generators, combined heat and power plants, 
photovoltaic panels, and wind turbines. Electricity is consumed by end-users and
electrical devices, e.g., water pumps and heat pumps. 
Active and reactive nodal power injections $P_{j, t}$ and $Q_{j, t}$ from bus $j$
at time $t$ are expressed as
\begin{gather}
	\begin{aligned}
	P_{j, t}=& \sum_{g \in \mathcal{G}_{j}} P_{g, t}^{\mathrm{G}}+
				\sum_{c \in \mathcal{C}_{j}} P_{c, t}^{\mathrm{C}}+
				\sum_{p \in \mathcal{P}_{j}} P_{p, t}^{\mathrm{PV}}+
				\sum_{w \in \mathcal{W}_{j}} P_{w, t}^{\mathrm{WT}}-\\
	& \sum_{(k, l) \in \mathcal{B}_{j}^{\mathrm{WP}}} P_{k l, t}^{\mathrm{WP}}-
		\sum_{h \in \mathcal{H}_{j}} P_{h, t}^{\mathrm{HP}}-
		P_{j, t}^{\mathrm{D}}, \forall j \in \mathcal{J},
	\end{aligned} \\
	\begin{aligned}
	Q_{j, t}=& \sum_{g \in \mathcal{G}_{j}} Q_{g, t}^{\mathrm{G}}+
				\sum_{c \in \mathcal{C}_{j}} Q_{c, t}^{\mathrm{C}}+
				\sum_{p \in \mathcal{P}_{j}} Q_{p, t}^{\mathrm{PV}}+\\
	& \sum\nolimits_{w \in \mathcal{W}_{j}} Q_{w, t}^{\mathrm{WT}}-
		Q_{j, t}^{\mathrm{D}}, \forall j \in \mathcal{J},
	\end{aligned}
\end{gather}
where $P_{g, t}^{\mathrm{G}}$, $P_{c, t}^{\mathrm{C}}$, $P_{p, t}^{\mathrm{PV}}$ 
and $P_{w, t}^{\mathrm{WT}}$ denote active power of the $g$-th generator and
the $c$-th CHP, and forecasted power from the photovoltaic array and the wind turbine \cite{wan2013probabilistic}, 
respectively; $P_{k l, t}^{\mathrm{WP}}$ and $P_{h, t}^{\mathrm{HP}}$ are 
active power consumed by the water pump located at pipe $(k,l)$ in DHS and
the $h$-th heat pump, respectively; 
symbols for reactive power are represented by $Q$ with corresponding superscripts
and subscripts; $\mathcal{J}$ refers to the set of buses in the electric power network;
$\mathcal{G}_{j}$, $\mathcal{C}_{j}$, $\mathcal{P}_{j}$, $\mathcal{W}_{j}$, and
$\mathcal{H}_{j}$ are sets of generators, CHP, photovoltaic panels, wind turbines,
and heat pumps that are connected to bus $j$, respectively; 
$\mathcal{B}_{j}^{\mathrm{WP}}$ is the set of pipes in DHS
with installation of water pumps that couple with bus $j$ in EPS.

The branch flow model is adopted to model power flow in radial 
distribution feeders. Nonlinear AC power flow models are transformed to tractable 
convex models using angle relaxation and conic relaxation 
\cite{farivar2013branch,low2014convex}.
Active and reactive power flow equations at bus $j$ in EPS are given as
\begin{gather}
	P_{j, t}=\sum_{k: j \rightarrow k} P_{j k, t}-
				\sum_{i: i \rightarrow j}\left(P_{i j, t}-r_{i j} l_{i j, t}\right), 
				\forall j \in \mathcal{J}, \\
	Q_{j, t}=\sum_{k: j \rightarrow k} Q_{j k, t}-
				\sum_{i: i \rightarrow j}\left(Q_{i j, t}-x_{i j} l_{i j, t}\right), 
				\forall j \in \mathcal{J},
\end{gather}
where $P_{i j, t}$ and $Q_{i j, t}$ are active and reactive power at bus $i$
flowing towards bus $j$ at time $t$, respectively; $r_{i j}$ and $x_{i j}$ are resistance and 
reactance of branch $(i,j)$, respectively; $l_{i j, t}$ refers to the squared 
magnitude of the branch current. 
Branch power flow defined by nodal voltages and currents are indicated as
\begin{gather}
	v_{j, t}\! =\! v_{i, t}\!-\! 2\left(\! r_{i j}\! P_{i j, t}\! +\! x_{i j}\! Q_{i j, t}\!\right)\! +\! 
				\left(r_{i j}^{2}\! +\! x_{i j}^{2}\right)\!  l_{i j, t},\!  \forall(i, j) \! \in\!  \mathcal{E}, \\
	\left\|
	\begin{array}{c}
	{2 P_{i j, t}},{2 Q_{i j, t}},{l_{i j, t}-v_{i, t}}
	\end{array}
	\right\|_{2} \leq l_{i j, t}+v_{i, t}, \forall(i, j) \in \mathcal{E},
\end{gather}
where $v_{i,t}$ denotes the squared voltage magnitude at bus $i$;
$\mathcal{E}$ refers to the set of branches in EPS. 
Boundary constraints on electric power generation and consumption are expressed as
\begin{gather}
	\underline{P}_{g}^{\mathrm{G}} \leq P_{g, t}^{\mathrm{G}} \leq \bar{P}_{g}^{\mathrm{G}}, 
	\underline{Q}_{g}^{\mathrm{G}} \leq Q_{g, t}^{\mathrm{G}} \leq \bar{Q}_{g}^{\mathrm{G}}, 
	\forall g \in \mathcal{G},\\
	\underline{P}_{c}^{\mathrm{C}} \leq P_{c, t}^{\mathrm{C}} \leq \bar{P}_{c}^{\mathrm{C}}, 
	\underline{Q}_{c}^{\mathrm{C}} \leq Q_{c, t}^{\mathrm{C}} \leq \bar{Q}_{c}^{\mathrm{C}}, 
	\forall c \in \mathcal{C},\\
	\underline{P}_{h}^{\mathrm{HP}} \leq P_{h, t}^{\mathrm{HP}} \leq \bar{P}_{h}^{\mathrm{HP}},
	\forall h \in \mathcal{H},\\
	P_{p, t}^{\mathrm{PV}} \leq \bar{P}_{p, t}^{\mathrm{PV}}, Q_{p, t}^{\mathrm{PV}}=0, 
	\forall p \in \mathcal{P},\\
	P_{w, t}^{\mathrm{WT}} \leq \bar{P}_{w, t}^{\mathrm{WT}}, 
	Q_{w, t}^{\mathrm{WT}}=\eta_{w}^{\mathrm{WT}} P_{w, t}^{\mathrm{WT}}, 
	\forall w \in \mathcal{W},
\end{gather}
where $[\underline{P}_{g}^{\mathrm{G}},\bar{P}_{g}^{\mathrm{G}}]$, 
$[\underline{P}_{c}^{\mathrm{C}},\bar{P}_{c}^{\mathrm{C}}]$ and 
$[\underline{P}_{h}^{\mathrm{HP}},\bar{P}_{h}^{\mathrm{HP}}]$ 
are active power bounds of generators, CHP and heat pumps, respectively;
$[\underline{Q}_{g}^{\mathrm{G}},\bar{Q}_{g}^{\mathrm{G}}]$ and 
$[\underline{Q}_{c}^{\mathrm{C}},\bar{Q}_{c}^{\mathrm{C}}]$ 
denote reactive power bounds of generators and CHP, respectively;
$\eta_{w}^{\mathrm{WT}}$ refers to the reactive-active power ratio of wind turbines;
$\mathcal{G}$, $\mathcal{C}$, $\mathcal{P}$, $\mathcal{W}$ and $\mathcal{H}$ 
are sets of generators, CHP, photovoltaic panels, wind turbines, and
heat pumps, respectively.
Ramping constraints are given as
\begin{gather}
	\left|\Delta P_{g, t}^{\mathrm{G}}\right| \leq \Delta \bar{P}_{g}^{\mathrm{G}},
	\left|\Delta Q_{g, t}^{\mathrm{G}}\right| \leq \Delta \bar{Q}_{g}^{\mathrm{G}},
	\forall g \in \mathcal{G},\\
	\left|\Delta P_{c, t}^{\mathrm{C}}\right| \leq \Delta \bar{P}_{c}^{\mathrm{C}},
	\left|\Delta Q_{c, t}^{\mathrm{C}}\right| \leq \Delta \bar{Q}_{c}^{\mathrm{C}}, 
	\forall c \in \mathcal{C},\\
	\left|\Delta P_{h, t}^{\mathrm{HP}}\right| \leq \Delta \bar{P}_{h}^{\mathrm{HP}}, 
	\forall h \in \mathcal{H},
\end{gather}
where $\Delta \bar{P}_{g, t}^{\mathrm{G}}$, $\Delta \bar{P}_{c, t}^{\mathrm{C}}$, 
and $\Delta \bar{P}_{h, t}^{\mathrm{HP}}$ are maximum ramping rates 
of active power for generators, CHP, heat pumps, respectively;
$\Delta \bar{Q}_{g, t}^{\mathrm{G}}$ and $\Delta \bar{Q}_{c, t}^{\mathrm{C}}$ are
reactive power ramping capability of generators and CHP, respectively.
Limits on voltages and currents are given as
\begin{equation}
	\underline{v}_{j} \leq v_{j, t} \leq \bar{v}_{j}, \forall j \in \mathcal{J}, 
	l_{i j, t} \leq \bar{l}_{i j}, \forall(i, j) \in \mathcal{E},
\end{equation}
where $\underline{v}_{j}$ and $\bar{v}_{j}$ are squared lower and upper bounds
of voltage magnitude; $\bar{l}_{i j}$ are squared maximum current magnitude. 

\subsection{District Heating System}

Modeling of DHS generally includes two sets of equations: hydraulic and thermal. 
Hydraulic equations focus on pressure conditions to determine the rate of heat delivery. 
Thermal equations characterize the changes in temperature levels within DHS.

Hydraulic dynamics are quickly transferred to the whole network within seconds,
while temperature dynamics are quite slow \cite{palsson1999equivalent}.
Thus, mass flow rates and pressure conditions are calculated based on
a static hydraulic model.
Temperature changes are characterized using the dynamic approach.

\subsubsection{Hydraulic Model}Water flow in networked pipelines is assumed to be incompressible.
Hence the distribution of mass flow is subject to continuity constraints,
where the sum of incoming mass flow equals to the total outflow
at each node, given as
\begin{equation}
	m_{k, t}^{\mathrm{in}}-m_{k, t}^{\mathrm{out}}=\sum_{l: k \rightarrow l} m_{k l, t}-
	\sum_{j: j \rightarrow k} m_{j k, t}, \forall k \in \mathcal{N},
\end{equation}
where $m_{k, t}^{\mathrm{in}}$ and $m_{k, t}^{\mathrm{out}}$ indicate 
mass flow injection and outflow at node $k$, respectively; 
$m_{k l, t}$ denotes the mass flow rate in pipe $(k,l)$ at time $t$,
and $m_{j k, t}$ represents the mass flow from node $j$ to node $k$;
$\mathcal{N}$ is the set of nodes in DHS.

Water is distributed from points of high pressure to points of low pressure. 
Major pressure losses occur when water travels through pipelines due to
friction against pipe walls.
Minor pressure losses are induced by turbulence through fittings and 
appurtenances.
The magnitude of major pressure losses is relevant to the water flow speed,
internal roughness and pipe dimension, formulated as
\begin{equation}
\label{equ:MajorPreLoss}
	p_{k, t}^{\mathrm{S}}-p_{l, t}^{\mathrm{S}}=\mu_{k l} m_{k l, t}^{2}, 
	p_{l, t}^{\mathrm{R}}-p_{k, t}^{\mathrm{R}}=\mu_{k l} m_{l k, t}^{2}, 
	\forall(k, l) \in \mathcal{B},
\end{equation}
where $p_{k, t}^{\mathrm{S}}$ and $p_{k, t}^{\mathrm{R}}$ are supply and return 
pressure at node $k$, respectively;
$\mathcal{B}$ is the set of pipes. 
The friction loss coefficient $\mu_{k l}$ can be derived based on
dimensional analysis \cite{walski2001water}, given as
\begin{equation}
	\mu_{k l}=f_{k l}^{\mathrm{D}} c_{\mathrm{Pa}}^{\mathrm{m}}
				\frac{8 L_{k l}}{D_{k l}^{5} \rho_{\mathrm{w}}^{2} \pi^{2} g_{a}},
	f_{k l}^{\mathrm{D}}=\frac{1.325}
				{\left[\ln \left(\frac{\epsilon}{3.7 D_{k l}}+
				\frac{5.74}{\operatorname{Re}^{0.9}}\right)\right]^{2}},
\end{equation}
where $f_{k l}^{\mathrm{D}}$ is the Darcy-Weisbach friction factor;
$c_{\mathrm{Pa}}^{\mathrm{m}}$ converts a pressure head 
(the height of water associated with its particular pressure, in units of meters)
to a pressure (Pa); 
$L_{k l}$ and $D_{k l}$ are the pipe length and diameter, respectively;
$\rho_{\mathrm{w}}$ denotes water density; 
$g_{a}$ is the gravitational acceleration constant;
$\epsilon$ is the pipe internal roughness; 
$\operatorname{Re}$ is the Reynolds number \cite{walski2001water}.
Minor pressure losses are mainly induced by bends and reducers of heat-exchangers.
Let $f_{k}^{\mathrm{m}}$ denote the minor loss coefficient, 
and $A_{k}$ be the cross-sectional area of heat-exchangers.
Minor pressure losses are given by
\begin{equation}
\label{equ:MinorPreLoss}
	p_{k, t}^{\mathrm{S}}-p_{k, t}^{\mathrm{R}} =
			\frac{f_{k}^{\mathrm{m}} c_{\mathrm{Pa}}^{\mathrm{m}}\left(m_{k, t}^{\text{out}}\right)^{2}} 
			{2 g_{a} \rho_{\mathrm{w}}^{2} A_{k}^{2}}, 
	\forall k \in \mathcal{N}.
\end{equation}

Valves are installed in DHS for flow control or pressure regulation.
Constraint (\ref{equ:Val}) denotes pressure differentials provided by
pressure-reducing valves at certain nodes, shown as 
\begin{equation}
	\label{equ:Val}
	\Delta p_{k l, t}^{\mathrm{VL}}=p_{k, t}^{\mathrm{S}}-p_{l, t}^{\mathrm{S}}, 
	\Delta p_{k l, t}^{\mathrm{VL}} \geq 0, \forall(k, l) \in \mathcal{B}^{\mathrm{VL}},
\end{equation}
where $\Delta p_{k l, t}^{\mathrm{VL}}$ is the pressure difference 
of the valve located at pipe $(k,l)$; 
$\mathcal{B}^{\mathrm{VL}}$ is the set of pipes with valves installed.

Water pumps raise potentials 
to overcome pressure losses in water distribution.
Variable-speed electric centrifugal pumps are modeled here
via the pump characteristic curve \cite{burgschweiger2009optimization}.
The relationship between the pressure difference and the flow rate
is given as
\begin{equation}
\label{equ:WatPump}
	p_{l, t}^{\mathrm{S}}\!- p_{k, t}^{\mathrm{S}}\!\! =\! c_{\mathrm{Pa}}^{\mathrm{m}} w_{k l, t}^{2}
	\!\!\left(\!\gamma_{k l}^{1}\!-\!\gamma_{k l}^{2}
	\!\left(\!\frac{m_{k l, t}}{\rho_{\mathrm{w}} w_{k l, t}}\!\right)^{\gamma_{k l}^{3}}\!\right)\!,\! 
	\forall(k,\! l) \!\in\! \mathcal{B}^{\mathrm{WP}}\!\!, 
\end{equation}
where $w_{k l, t}$ indicates the relative pump speed;
$\gamma_{k l}^{1}$, $\gamma_{k l}^{2}$, and $\gamma_{k l}^{3}$ are
coefficients estimated through empirical data; 
$\mathcal{B}^{\mathrm{WP}}$ is the set of pipes with water pumps.
Pump electric power is determined by
\begin{equation}
\label{equ:WatPumpPow}
	P_{k l, t}^{\mathrm{WP}}\!=\!
	\frac{m_{k l, t}\left(p_{l, t}^{\mathrm{S}}\!-\!p_{k, t}^{\mathrm{S}}\right)}
			{\eta_{k l}^{\mathrm{WP}} \rho_{\mathrm{w}}}, 
	P_{k l, t}^{\mathrm{WP}} \!\leq\! \bar{P}_{k l}^{\mathrm{WP}}, 
	\forall(k, l) \!\in\! \mathcal{B}^{\mathrm{WP}},
\end{equation}
where $\eta_{k l}^{\mathrm{WP}}$ is the pump efficiency; 
$\bar{P}_{k l}^{\mathrm{WP}}$ denotes the electric power limit of water pumps.
Constraints on nodal pressure and mass flow rates are indicated as
\begin{gather}
	\underline{p}_{k}^{\mathrm{S}} \leq p_{k, t}^{\mathrm{S}} \leq \bar{p}_{k}^{\mathrm{S}},
	\underline{p}_{k}^{\mathrm{R}} \leq p_{k, t}^{\mathrm{R}} \leq \bar{p}_{k}^{\mathrm{R}},
	\forall k \in \mathcal{N}, \\
	\underline{m}_{k l} \leq m_{k l, t} \leq \bar{m}_{k l},
	\forall(k, l) \in \mathcal{B},
\end{gather}
\begin{gather}
	\underline{m}_{k} \leq m^\mathrm{in}_{k,t}, m^\mathrm{out}_{k,t} 
	\leq \bar{m}_{k},
	\forall k \in \mathcal{N},
\end{gather}
where $[\underline{p}_{k}^{\mathrm{S}},\bar{p}_{k}^{\mathrm{S}}]$ and 
$[\underline{p}_{k}^{\mathrm{R}},\bar{p}_{k}^{\mathrm{R}}]$ 
denote pressure bounds in supply and return networks;
$[\underline{m}_{k l},\bar{m}_{k l}]$ and $[\underline{m}_{k},\bar{m}_{k}]$ 
are mass flow limits.

\subsubsection{Thermal Model}

Generated heat is shown as 
\begin{equation}
	H_{c, t}^{\mathrm{C}}=\eta_{c}^{\mathrm{C}} P_{c, t}^{\mathrm{C}}, 
	\forall c \in \mathcal{C}, 
	H_{h, t}^{\mathrm{HP}}=\eta_{h}^{\mathrm{HP}} P_{h, t}^{\mathrm{HP}}, 
	\forall h \in \mathcal{H},
\end{equation}
where $H_{c, t}^{\mathrm{C}}$ and $H_{h, t}^{\mathrm{HP}}$ are heat produced by
the $c$-th CHP and the $h$-th heat pump, respectively;
$\eta_{c}^{\mathrm{C}}$ and $\eta_{h}^{\mathrm{HP}}$ represent
the power-to-heat ratio of CHP and the coefficient of performance of a heat pump,
respectively.
Total heat is transferred to the heating network
at heat sources and delivered to end-users at heat-exchangers,
determined by the product of mass flow rates and temperatures rises (or drops).
Heat flow at production and user sites satisfy
\begin{gather}
\label{equ:HeatTransSource}
	\sum_{c \in \mathcal{C}_{k}}\! H_{c, t}^{\mathrm{C}}\!+\!
	\sum_{h \in \mathcal{H}_{k}}\! H_{h, t}^{\mathrm{HP}}\!=\!
	c_{\mathrm{w}} m_{k, t}^{\mathrm{in}}
	\!\left(\!T_{k, t}^{\mathrm{S}, \mathrm{in}}-T_{k, t}^{\mathrm{R}}\!\right)\!, 
	\forall k \in \mathcal{N}, \\
\label{equ:HeatTransLoad}
	H_{k, t}^{\mathrm{D}}=c_{\mathrm{w}} m_{k, t}^{\mathrm{out}}
	\left(T_{k, t}^{\mathrm{S}}-T_{k, t}^{\mathrm{R}, \mathrm{out}}\right), 
	\forall k \in \mathcal{N},
\end{gather}
where $T_{k, t}^{\mathrm{S}, \mathrm{in}}$ and $T_{k, t}^{\mathrm{R}}$ are 
the supply temperature from the heat source at node $k$ and 
the nodal temperature at node $k$ in the return heating network, respectively;
$T_{k, t}^{\mathrm{S}}$ and $T_{k, t}^{\mathrm{R}, \mathrm{out}}$ denote 
the nodal temperatures at node $k$ in the supply network and 
the return temperatures from the heat user at node $k$, respectively;
$H_{k, t}^{\mathrm{D}}$ refers to the heat load at node $k$ in DHS;
$c_{\mathrm{w}}$ is the specific heat capacity of water; 
$\mathcal{C}_{k}$ and $\mathcal{H}_{k}$ are sets of CHP and heat pumps at node $k$.

We assume that all incoming water flow is fully mixed at each node.
According to the law of energy conservation, 
the nodal temperature after mixing all incoming heating flow in the supply and
return heating networks is determined by
\begin{gather}
\label{equ:TempMixSup}
	T_{l, t}^{\mathrm{S}}=
	\frac{T_{l, t}^{\mathrm{S,in}} m_{l, t}^{\mathrm{in}}+
			\sum_{k: k \rightarrow l} T_{k l, t}^{\mathrm{S,out}} m_{k l, t}}
	{m_{l, t}^{\mathrm{in}}+\sum_{k: k \rightarrow l} m_{k l, t}}, 
	\forall l \in \mathcal{N}, \\
\label{equ:TempMixRet}
	T_{k, t}^{\mathrm{R}}=
	\frac{T_{k, t}^{\mathrm{R,out}} m_{k, t}^{\mathrm{out}}+
			\sum_{l: l \rightarrow k} T_{l k, t}^{\mathrm{R,out}} m_{l k, t}}
	{m_{k, t}^{\mathrm{out}}+\sum_{l: l \rightarrow k} m_{l k, t}}, 
	\forall k \in \mathcal{N},
\end{gather}
where $T_{k l, t}^{\mathrm{S,out}}$ and $T_{l k, t}^{\mathrm{R,out}}$ 
are the outlet temperatures of pipe $(k,l)$ in supply and return networks, respectively.
The correlation between pipeline inlet and nodal temperatures
is given by
\begin{equation}
	T_{k l, t}^{\mathrm{S,in}}=T_{k, t}^{\mathrm{S}},
	T_{l k, t}^{\mathrm{R,in}}=T_{l, t}^{\mathrm{R}},
	\forall(k, l) \in \mathcal{B},
\end{equation}
where $T_{k l, t}^{\mathrm{S,in}}$ and $T_{l k, t}^{\mathrm{R,in}}$ 
are inlet temperature of pipe $(k,l)$ in supply and return networks, respectively.

Time delays of heat distribution are modeled by the node method 
\cite{palsson1999equivalent}.
In brief, the outlet temperature of a pipeline is defined as 
a nonlinear function of the inlet temperature at multiple previous time slots.
The loss-free outlet temperature $\dot{T}_{k l, t}^{\mathrm{S,out}}$
of pipe $(k,l)$ in the supply network is formulated as
\begin{equation}
\label{equ:LossFreeTemp}
	\begin{aligned}
	\dot{T}_{k l, t}^{\mathrm{S,out}}= &
	\Big[(M_{k l, t}^{\mathrm{st}}-M_{k l}) T_{k l, t-t_{k l, t}^{\mathrm{st}}}^{\mathrm{S,in}}+
			\Delta_{k l, t}^{\mathrm{S}}+ \\
			(m_{k l, t} &\Delta t+M_{k l}-M_{k l, t}^{\mathrm{ed}}) 
			T_{k l, t-t_{k l, t}^{\mathrm{ed}}}^{\mathrm{S,in}}\Big] 
	/\left(m_{k l, t} \Delta t\right),
	\end{aligned}
\end{equation}
where $\Delta t$ is the length of time step; 
time delays $t_{k l, t}^{\mathrm{st}}$ and $t_{k l, t}^{\mathrm{ed}}$ 
basically denote heat delivery time for hot water 
traveling through pipe $(k,l)$ at time $t$;
$t_{k l, t}^{\mathrm{st}}$ can be regarded as the ``arriving'' time
delay, while $t_{k l, t}^{\mathrm{ed}}$ is the ``leaving'' time delay, indicated as
\begin{gather}
\label{equ:TimeDelaySt}
	t_{k l, t}^{\mathrm{st}} = \min
	\Bigg\{\zeta\geq 0; \sum_{\tau=t-\zeta}^t m_{k l, \tau} \Delta t>M_{k l}\Bigg\},\\
\label{equ:TimeDelayEd}
	t_{k l, t}^{\mathrm{ed}} = \min
	\Bigg\{\zeta\geq 0; \sum_{\tau=t-\zeta}^t m_{k l, \tau} \Delta t>M_{k l} + 
			m_{k l, t} \Delta t\Bigg\}.
\end{gather}
Other auxiliary variables are formulated as
\begin{gather}
	M_{k l}=\frac{\pi L_{k l} D_{k l}^{2} \rho_{\mathrm{w}}}{4},
	M_{k l, t}^{\mathrm{st}}=\sum_{\tau=t-t_{k l, t}^{\mathrm{st}}}^{t} 
								m_{k l, \tau} \Delta t,\\
	M_{k l, t}^{\mathrm{ed}}=\left\{\begin{aligned}
	&\sum_{\tau=t-t_{k, t}^{\mathrm{ed}}+1}^{t} m_{k l, \tau} 
		\Delta t, t_{k l, t}^{\mathrm{ed}}>t_{kl,t}^{\mathrm{st}} \\
	&M_{k l, t}^{\mathrm{st}}, t_{k l, t}^{\mathrm{ed}}=t_{k l, t}^{\mathrm{st}}
	\end{aligned}\right.,
\end{gather}

\begin{equation}
	\Delta_{k l, t}^{\mathrm{S}}=\sum_{\tau=t-t_{k l, t}^{\mathrm{ed}}+1}^{t-t_{k l, t}^{\mathrm{st}}-1} 
					m_{k l, \tau} \Delta t T_{k l, \tau}^{\mathrm{S,in}},
\end{equation}
where $M_{k l}$ refers to the total mass of water in pipe $(k,l)$;
$M_{k l, t}^{\mathrm{st}}$ and $M_{k l, t}^{\mathrm{ed}}$ are 
summations of water mass at different time instants.
Considering heat losses appearing in water distribution,
the outlet temperature $T_{k l, t}^{\mathrm{S,out}}$
of pipe $(k,l)$ at time $t$ is finally formulated as 
\begin{equation}
\label{equ:RealTemp}
	\begin{aligned}
	T_{k l, t}^{\mathrm{S,out}}\!\!=&T_{t}^{\mathrm{GD}}
	+\left(\dot{T}_{kl,t}^{\mathrm{S,out}}-T_{t}^{\mathrm{GD}}\right) \times \\
	& \text{exp}\left\{ \!-\!\frac{\lambda_{kl} \Delta t}{A_{k l} \rho_{\mathrm{w}} c_{\mathrm{w}}}
	\left(t_{kl,t}^{\mathrm{st}}\!+\!\frac{1}{2}\!+\!
	\frac{ M_{kl,t}^{\mathrm{ed}}-M_{kl,t}^{\mathrm{st}} }
			{ m_{kl,t-t_{kl,t}^{\mathrm{st}}} \Delta t} \right) \right\},
	\end{aligned}
\end{equation}
where $T_{t}^{\mathrm{GD}}$ refers to the ground temperature;
$\lambda_{kl}$ denotes the heat conductivity (W/m/K) of pipe $(k,l)$ in DHS.
The heat distribution in the return network can be derived similarly. 
A visualized illustration can be found in \cite{li2015combined}. 

Network constraints on nodal temperature in supply and return networks 
are denoted as
\begin{equation}
	\underline{T}_{k}^{\mathrm{S}} \leq T_{k, t}^{\mathrm{S}},T_{k, t}^{\mathrm{S,in}} 
	\leq \bar{T}_{k}^{\mathrm{S}}, 
	\underline{T}_{k}^{\mathrm{R}} \leq T_{k, t}^{\mathrm{R}},T_{k, t}^{\mathrm{R,out}} 
	\leq \bar{T}_{k}^{\mathrm{R}}, 
	\forall k \in \mathcal{N},
\end{equation}
where $[\underline{T}_{k}^{\mathrm{S}},\bar{T}_{k}^{\mathrm{S}}]$ and 
$[\underline{T}_{k}^{\mathrm{R}},\bar{T}_{k}^{\mathrm{R}}]$ 
are nodal temperature bounds in supply and return networks, respectively;

\section{Simplification and Relaxation}

\subsection{Simplified Thermal Dynamic Model}

As detailed from (\ref{equ:LossFreeTemp}) to (\ref{equ:RealTemp}), 
the node method uses complex nonlinear equations 
(e.g., exponential terms in (\ref{equ:RealTemp}))
and logical conditions 
(e.g., time delays (\ref{equ:TimeDelaySt}) and (\ref{equ:TimeDelayEd}))
to characterize temperature dynamics and heat losses in DHS.
The node method is accurate and easy to understand from the modeling perspective.
However, it is computationally not affordable in optimization.
For instance, the loss-free outlet temperature $\dot{T}_{k l, t}^{\mathrm{S,out}}$
is mapped with the inlet temperature $T_{kl,\tau}^{\mathrm{S,in}}$
at multiple time instants from $\tau=t-t^\mathrm{ed}_{kl,t}$ to 
$\tau=t-t^\mathrm{st}_{kl,t}$. 
Note that $t^\mathrm{st}_{kl,t}$ and $t^\mathrm{ed}_{kl,t}$ 
are both decision variables satisfying logic conditions in (\ref{equ:TimeDelaySt}) and (\ref{equ:TimeDelayEd}),
and thus the temperature change equation (\ref{equ:LossFreeTemp}) is ``indeterminate''.
To reformulate (\ref{equ:LossFreeTemp})-(\ref{equ:TimeDelayEd}) as ``exact'' constraints in optimization,
auxiliary integer variables are introduced based on the Big-M method 
\cite{mitridati2018power}. 
The numbers of additional integer variables are 
proportional to the maximum time delay, the number of pipelines, and time periods.

To avoid the combinatorial inefficiency induced by 
logical constraints and auxiliary integer variables, 
the STD model is derived based on DHS dynamics.
The partial differential equation 
capturing influence of flow speed and heat convection is given as 
\cite{palsson1999equivalent}
\begin{equation}
	\frac{\partial T}{\partial t}+
	\frac{m}{A \rho_\mathrm{w}} \frac{\partial T}{\partial x}+
	\frac{\lambda}{A \rho_\mathrm{w} c_\mathrm{w}}\left(T-T^\mathrm{GD}\right)=0,
\end{equation}
where $T$ and $m$ denote temperature and mass flow rate, respectively;
$T^\mathrm{GD}$ is the temperature of the surrounding ground;
$\lambda$ refers to the thermal transfer coefficient (W/m/K).
The partial derivatives are approximated using a first-order upwind scheme
with respect to time $t$ and distance $x$. 
This discretization is applied to each pipeline at each time instant.
Eventually, temperature dynamics in the supply heating network are modeled as 
\begin{equation}
\label{equ:STDSup}
	\begin{aligned}
	\frac{T_{kl,t}^{\mathrm{S}}-T_{kl,t-1}^{\mathrm{S}}}{\Delta t}+&
	\frac{m_{kl,t}}{A_{kl} \rho_{\mathrm{w}}}
	\frac{T_{kl,t}^{\mathrm{S,out}}-T_{kl,t}^{\mathrm{S,in}}}{L_{k l}}+\\
	& \frac{\lambda_{kl}}{A_{kl} \rho_{\mathrm{w}} c_{\mathrm{w}}}
	\left(T_{kl,t}^{\mathrm{S}}-T_{t}^{\mathrm{GD}}\right)=0,
	\end{aligned}
\end{equation}
where $T_{kl,t}^{\mathrm{S}}=(T_{kl,t}^{\mathrm{S,in}}+T_{kl,t}^{\mathrm{S,out}})/2$
refers to the average supply temperature of pipe $(k,l)$ at time $t$.
Similarly, temperature dynamics in the return network are given by 
\begin{equation}
	\label{equ:STDRet}
		\begin{aligned}
		\frac{T_{lk,t}^{\mathrm{R}}-T_{lk,t-1}^{\mathrm{R}}}{\Delta t}+&
		\frac{m_{lk,t}}{A_{kl} \rho_{\mathrm{w}}}
		\frac{T_{lk,t}^{\mathrm{R,out}}-T_{lk,t}^{\mathrm{R,in}}}{L_{k l}}+\\
		& \frac{\lambda_{kl}}{A_{kl} \rho_{\mathrm{w}} c_{\mathrm{w}}}
		\left(T_{lk,t}^{\mathrm{R}}-T_{t}^{\mathrm{GD}}\right)=0,
	\end{aligned}
\end{equation}
where $T_{lk,t}^{\mathrm{R}}=(T_{lk,t}^{\mathrm{R,out}}+T_{lk,t}^{\mathrm{R,in}})/2$ 
is the average return temperature of pipe $(k,l)$ at time $t$.
The proposed STD model accounts for heat storage of pipes, 
heat losses to surroundings and time delays of heat distribution.
Two set of equations (\ref{equ:STDSup}) and (\ref{equ:STDRet}) are 
incorporated into the CHPD model instead of (\ref{equ:LossFreeTemp})-(\ref{equ:RealTemp})
to enable a computationally efficient model.

\subsection{Conic and Polyhedral Relaxation}

Nonconvex parts in the CHPD model mainly come from quadratic equality 
constraints and bilinear constraints.
Quadratic equality constraints include 
pressure losses (\ref{equ:MajorPreLoss}) and (\ref{equ:MinorPreLoss}).
Bilinear constraints are used in pump power equations (\ref{equ:WatPumpPow}),
heat transfer equations (\ref{equ:HeatTransSource}) and (\ref{equ:HeatTransLoad}),
temperature mixing equations (\ref{equ:TempMixSup}) and (\ref{equ:TempMixRet}),
and the STD model (\ref{equ:STDSup}) and (\ref{equ:STDRet}).
The convex CHPD model is derived based on conic and polyhedral relaxations
to form a second-order cone programming (SOCP) problem.

Quadratic equality constraints are reformulated as second-order cone constraints
by replacing equal signs with inequality signs. 
Thus, major and minor pressure losses are reformulated as 
\begin{equation}
	\label{equ:MajorPreLossRelax}
		p_{k, t}^{\mathrm{S}}-p_{l, t}^{\mathrm{S}}\geq\mu_{k l} m_{k l, t}^{2}, 
		p_{l, t}^{\mathrm{R}}-p_{k, t}^{\mathrm{R}}\geq\mu_{k l} m_{l k, t}^{2}, 
		\forall(k, l) \in \mathcal{B},
\end{equation}
\begin{equation}
	\label{equ:MinorPreLossRelax}
		p_{k, t}^{\mathrm{S}}-p_{k, t}^{\mathrm{R}} \geq
				\frac{f_{k}^{\mathrm{m}} c_{\mathrm{Pa}}^{\mathrm{m}}\left(m_{k, t}^{\text{out}}\right)^{2}} 
				{2 g_{a} \rho_{\mathrm{w}}^{2} A_{k}^{2}}, 
		\forall k \in \mathcal{N}.
\end{equation}

Bilinear constraints are relaxed based on polyhedral relaxation. 
Polyhedral relaxation provides an outer convex hull for bilinear terms based 
on McCormick envelopes, which retain linearity and minimizes the size
of the relaxed feasible space.
For instance, the heat transfer function (\ref{equ:HeatTransSource}) is 
reformulated as
\begin{gather}
\label{equ:BLOne}
	\begin{aligned}
	H^\mathrm{D}_{k,t} \geq & 
	c_\mathrm{w} \underline{m}_{k} (T^\mathrm{S}_{k,t}-T^\mathrm{R,out}_{k,t}) +
	c_\mathrm{w} m^\mathrm{in}_{k,t} (\underline{T}^\mathrm{S}_{k}-\bar{T}^\mathrm{R}_k) \\
	& - c_\mathrm{w} \underline{m}_{k} (\underline{T}^\mathrm{S}_{k}-\bar{T}^\mathrm{R}_k),
	\end{aligned} \\
	\begin{aligned}
	H^\mathrm{D}_{k,t} \geq & 
	c_\mathrm{w} \bar{m}_{k} (T^\mathrm{S}_{k,t}-T^\mathrm{R,out}_{k,t}) +
	c_\mathrm{w} m^\mathrm{in}_{k,t} (\bar{T}^\mathrm{S}_{k}-\underline{T}^\mathrm{R}_k) \\
	& - c_\mathrm{w} \bar{m}_{k} (\bar{T}^\mathrm{S}_{k}-\underline{T}^\mathrm{R}_k),
	\end{aligned} \\
	\begin{aligned}
	H^\mathrm{D}_{k,t} \leq & 
	c_\mathrm{w} \underline{m}_{k} (T^\mathrm{S}_{k,t}-T^\mathrm{R,out}_{k,t}) +
	c_\mathrm{w} m^\mathrm{in}_{k,t} (\bar{T}^\mathrm{S}_{k}-\underline{T}^\mathrm{R}_k) \\
	& - c_\mathrm{w} \underline{m}_{k} (\bar{T}^\mathrm{S}_{k}-\underline{T}^\mathrm{R}_k),
	\end{aligned} \\
\label{equ:BLFour}
	\begin{aligned}
	H^\mathrm{D}_{k,t} \geq & 
	c_\mathrm{w} \bar{m}_{k} (T^\mathrm{S}_{k,t}-T^\mathrm{R,out}_{k,t}) +
	c_\mathrm{w} m^\mathrm{in}_{k,t} (\underline{}{T}^\mathrm{S}_{k}-\bar{T}^\mathrm{R}_k) \\
	& - c_\mathrm{w} \bar{m}_{k} (\underline{}{T}^\mathrm{S}_{k}-\bar{T}^\mathrm{R}_k),
	\end{aligned}
\end{gather}
where the bilinear constraint (\ref{equ:HeatTransSource}) is replaced by four linear inequality
constraints (\ref{equ:BLOne})-(\ref{equ:BLFour}). 
Polyhedral constraints for other bilinear constraints can be derived similarly.

\begin{figure}[htb]
	\centering
	\includegraphics[width=0.80\columnwidth]{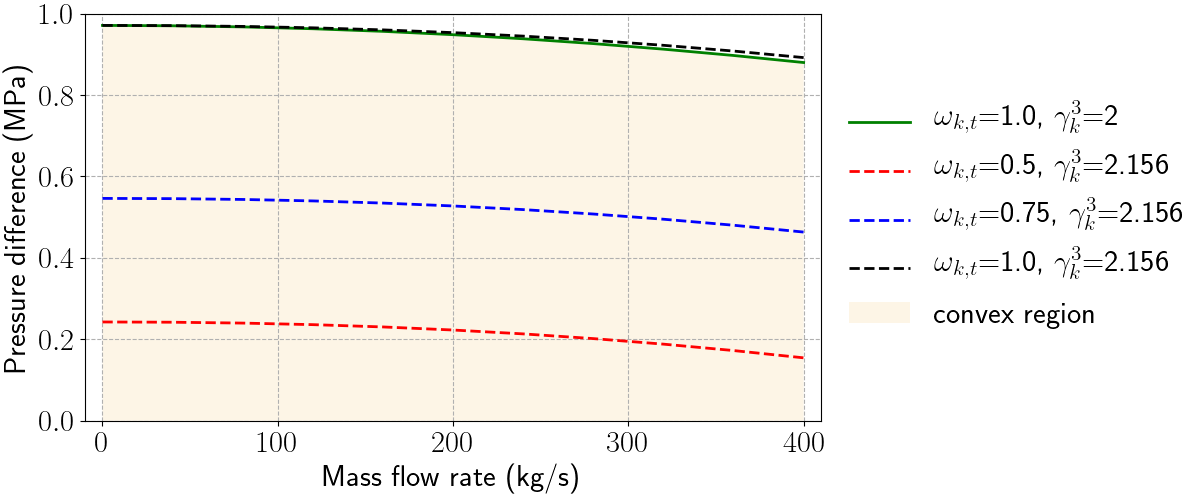}
	\caption{Convex operating region of a water pump.}
	\label{fig:ConvexWatPump}
\end{figure}

In addition, the pump characteristic curve denoted as (\ref{equ:WatPump})
is also nonconvex. However, the coefficient $\gamma^3_{kl}$ usually takes 
a value around 2 \cite{burgschweiger2009optimization}. 
Thus, a convex operating region can be formulated
for a water pump using quadratic inequality constraints.
The relative pump speed $w_{kl,t}$ is eliminated by examining all possible values
to form the operating region. 
Fig. \ref{fig:ConvexWatPump} shows the convex operating region of a water pump with 
the $\gamma^1=99.02, \gamma^2=57.74, \gamma^3=2.156$. 
Red, blue and black dashed lines depict the pump characteristic curves 
when $w$ equals to 0.5, 0.75 and 1.0, respectively. 
The shaded area indicates the convex operating region of water pumps, expressed as
\begin{equation}
	p_{l, t}^{\mathrm{S}}-p_{k, t}^{\mathrm{S}} \leq
	c_{\mathrm{Pa}}^{\mathrm{m}} \gamma_{k l}^{1}-
	c_{\mathrm{Pa}}^{\mathrm{m}} \gamma_{k l}^{2}
			\frac{m_{k l, t}^2}{\rho_{\mathrm{w}}^2} , 
	\forall(k, l) \in \mathcal{B}^{\mathrm{WP}}.
\end{equation}

\section{Relaxation Tightening}

\subsection{Piecewise Polyhedral Relaxation}

\begin{figure}[htb]
	\centering
	\includegraphics[width=0.90\columnwidth]{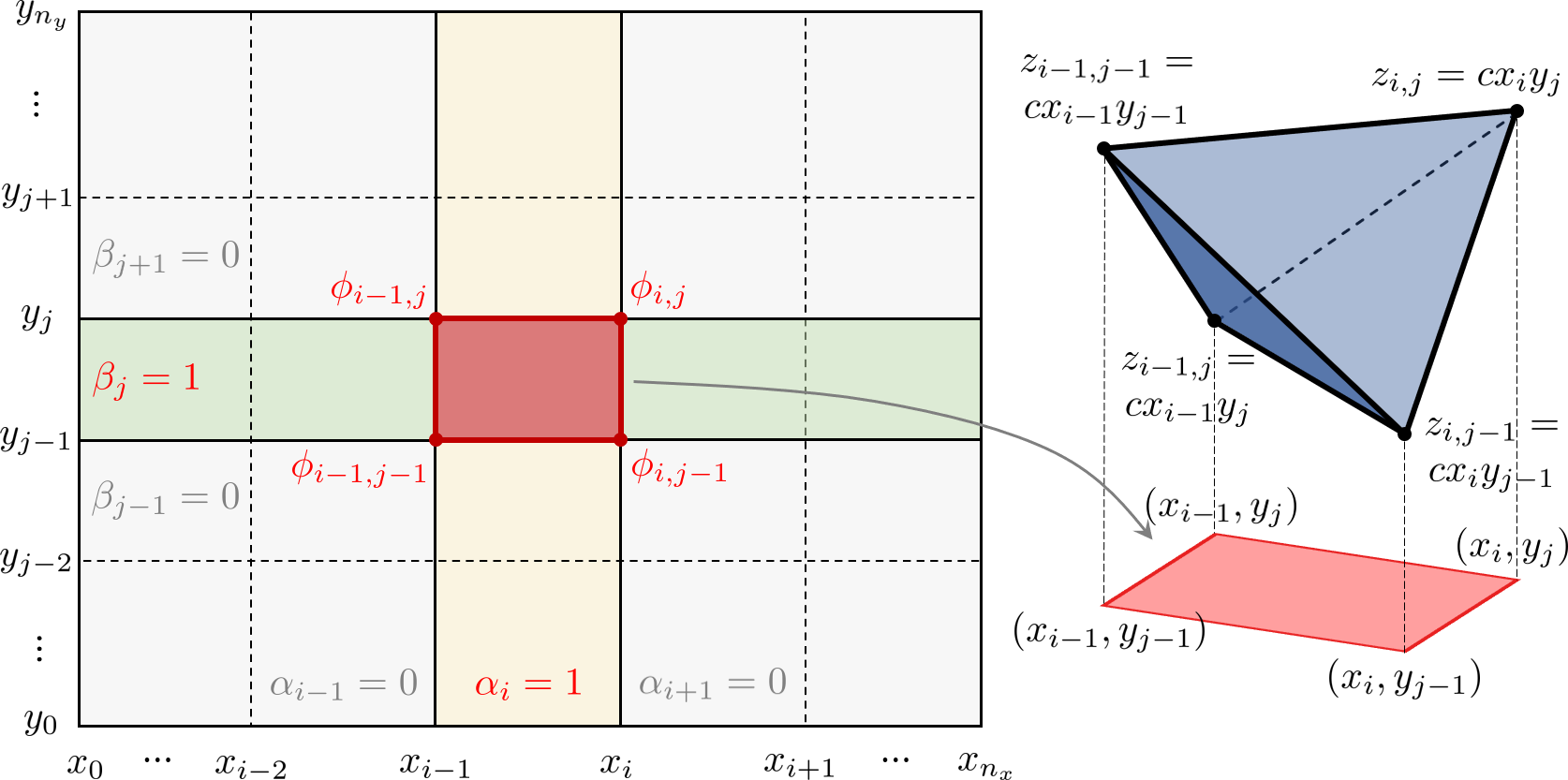}
	\caption{Illustration of piecewise polyhedral relaxation.}
	\label{fig:piecewise}
\end{figure}

The polyhedral relaxation based on McCormick envelopes is not tight
if participating variables in bilinear terms are all continuous \cite{mccormick1976computability}. 
To improve the relaxation quality, McCormick envelopes are tightened via
piecewise polyhedral relaxation. 
For the sake of brevity, the formulation for 
a general bilinear constraint $z=cxy$ with
$(x,y)\in[\underline{x},\bar{x}]\times[\underline{y},\bar{y}]$ is given here, 
where $z$, $x$ and $y$ are decision variables, and $c$ refers to a constant.
Without loss of generality,
the domains of $x$ and $y$ are split into $n$ disjoint regions 
with identical segment lengths. Then, $2n$ binary variables are introduced,
$\alpha_i, \beta_i, i\in\mathbb{I}_1^n$, 
to indicate which partitions in $x$ domain and $y$ domain are active. 
$\mathbb{I}_a^b$ refers to the set of integer variables from $a$ to $b$.
Lower and upper bounds of $x$ and $y$ on the $i$-th partition ($i \in \mathbb{I}_1^n$)
are calculated by 
\begin{equation}
\label{equ:partitionx}
	x_{i-1} = \underline{x} + (i-1)(\bar{x}-\underline{x})/n, 
	x_i = \underline{x} + i(\bar{x}-\underline{x})/n, 
\end{equation}
\begin{equation}
\label{equ:partitiony}
	y_{i-1} = \underline{y} + (i-1)(\bar{y}-\underline{y})/n, 
	y_i = \underline{y} + i(\bar{y}-\underline{y})/n, 
\end{equation}
where $x_{i-1}$ and $x_i$ are lower and upper bounds of $x$ 
on the $i$-th partition; similarly, 
$y_{i-1}$ and $y_i$ are lower and upper bounds of $y$ 
on the $i$-th partition. A convex combination is used to formulate a
piecewise polyhedral relaxation of bilinear equations, given as 
\begin{equation}
	x \!=\! \sum_{i=0}^n\! \sum_{j=0}^n \!\phi_{i,j} x_i, 
	y \!=\! \sum_{i=0}^n\! \sum_{j=0}^n \!\phi_{i,j} y_j,
	z \!=\! \sum_{i=0}^n\! \sum_{j=0}^n \!\phi_{i,j} c x_i y_j,
\end{equation}
\begin{equation}
	0 \leq \phi_{i,j} \leq 1, \forall i \in \mathbb{I}_0^n,
	\forall j \in \mathbb{I}_0^n,
\end{equation}
\begin{equation}
	\sum_{i=1}^n \alpha_i = 1, 
	\sum_{i=1}^n \beta_i = 1,
	\sum_{i=0}^n \sum_{j=0}^n \phi_{i,j} = 1,
\end{equation}
\begin{equation}
	\phi_{0,j} \!\leq\! \alpha_1,
	\phi_{i,j} \!\leq\! \alpha_i+\alpha_{i+1}, \forall i \in \mathbb{I}_1^{n-1},
	\phi_{n,j} \!\leq\! \alpha_n, 
	\forall j \in \mathbb{I}_0^n,
\end{equation}
\begin{equation}
\label{equ:SOS}
	\phi_{i,0} \leq \beta_1,
	\phi_{i,j} \leq \beta_i+\beta_{i+1}, \forall j \in \mathbb{I}_1^{n-1},
	\phi_{i,n} \leq \beta_n, 
	\forall i \in \mathbb{I}_0^n,
\end{equation}
where $\phi_{i,j}$ is a continuous variable associated with the point of
$(x_i,y_j,z_{i,j}=cx_iy_j)$; 
binary variables $\alpha_i$ and $\beta_j$ are set to one if the $i$-th partition 
on the $x$ domain and the $j$-th partition on the $y$ domain are active. 
A detailed graphical illustration for the piecewise convex relaxation is 
given in Fig. \ref{fig:piecewise},
where the left-hand red box represents the active partition 
with $\alpha_i=1$ and $\beta_j=1$, namely,
$x_{i-1}\leq x\leq x_{i}$ and $y_{j-1}\leq y\leq y_{j}$.
The right side of Fig. \ref{fig:piecewise} shows a convex polyhedral envelope
for this particular partition.

\subsection{Adaptive Solution Algorithm}

The piecewise polyhedral relaxation outlined above is introduced to 
tighten the McCormick envelopes
based on mixed-integer second-order cone programming (MISOCP).
In general, the more integer variables, 
the better the relaxation is due to a tighter outer approximation
of the original feasible set. 
Conventional piecewise McCormick envelopes use uniform partitioning 
where piecewise relaxation occurs on all variable domains,
leading to MISOCP problems with many binary variables.
To promote a better relaxation without imposing a heavy computational burden,
an adaptive solution algorithm is proposed based on dynamic bivariate partitioning
to shrink the relaxed feasible region successively 
while balancing the computational requirement.

The main idea of the proposed adaptive strategy is 
to identify constraints with the highest violation and 
iteratively refine the domain partitions accordingly.
Consequently, a sequence of MISOCP problems is solved 
based on a successively tighter piecewise polyhedral relaxation.
The pseudocode for the adpative solution algorithm is reported in Algorithm \ref{alg:adaptive}.
Suppose that the CHPD model has $m$ bilinear constraints, which can be generally 
formulated as $z_i=c_i x_iy_i, i\in\mathbb{I}_1^m$. 
Each bilinear constraint is relaxed with $n_i$ partitions.
The termination criterion is time limit $\bar{t}$.
\begin{algorithm}[htb]
	\footnotesize
	\caption{The adaptive solution algorithm}
	\label{alg:adaptive}
	\begin{algorithmic}[1]
		\State \textbf{Initialize} $\bar{t}$ \Comment{time limit}
		\State $n_i\gets 1, \forall i\in\mathbb{I}_1^m$ \Comment{numbers of partitions}
		\State $PreviousSolution\gets$ NULL \Comment{initialize a previous solution}
		\State $Tstart,Tend\gets CurrentTime$
		\While{$Tend-Tstart\leq\bar{t}$}
			\State Implement piecewise relaxation based on 
			(\ref{equ:partitionx})-(\ref{equ:SOS}) and $n_i, i\in\mathbb{I}_1^m$
			\label{pos:chpd}
			\State Formulate a relaxed CHPD-MISOCP model 
			\State Import $PreviousSolution$ as a warm start
			\State Solve the problem using Gurobi
			\State $PreviousSolution\gets CurrentSolution$ \Comment{save results}
			\State Get solutions of bilinear terms as $x^\ast_i,y^\ast_i,z^\ast_i,i\in\mathbb{I}_1^m$
			\State $\bar{gap}\gets 0$, $I\gets 1$
			\For{$i\in\mathbb{I}_1^m$}
				\State $gap_i\gets \frac{z^\ast_i-c_ix^\ast_iy^\ast_i}{z^\ast_i}\times 100\%$ 
				\Comment{constraint violation rate}
				\If{$gap_i>\bar{gap}$}
					\State $\bar{gap}\gets gap_i$
					\State $I\gets i$
				\EndIf
			\EndFor
			\State $n_I\gets n_I+1$ \Comment{refine partitions}
			\State $Tend\gets CurrentTime$ \Comment{refresh run time}
		\EndWhile	
	\end{algorithmic}
\end{algorithm}

\begin{figure}[htb]
	\centering
	\includegraphics[width=\columnwidth]{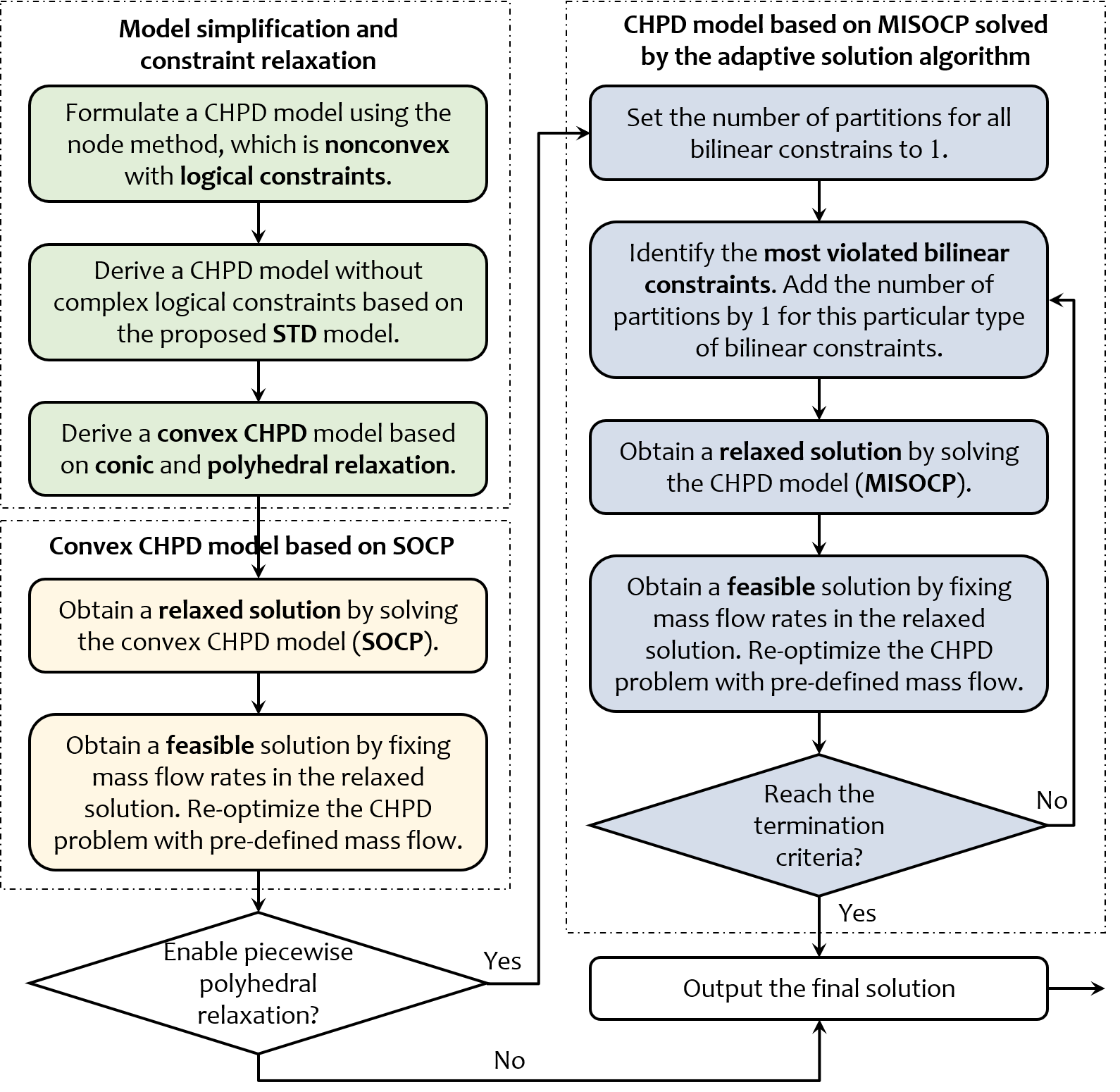}
	\caption{Flowchart of the solution strategy for CHPD.}
	\label{fig:adaptive}
\end{figure}

Fig. \ref{fig:adaptive} illustrates the main steps in the overall
solution strategy. Model simplification and constraint relaxation are 
conducted first to derive a convex CHPD model. 
Then, two choices are offered depending on the operator preference. 
The reason is that 
integer variables generally increase computing efforts
due to the exponential complexity for branching and bounding.
Besides, solvers for MISOCP are also not 
as scalable as those for mixed-integer linear programming.

If system operators aim to find a feasible solution 
with limited time or computing resources, the convex CHPD model based on 
SOCP is solved to obtain a relaxed solution.
A feasible solution with respect to the original CHPD problem is then
recovered by fixing mass flow rates and re-optimizing the CHPD problem with 
known hydraulic conditions. 
On the other hand, if the solution optimality is the first concern instead of 
computing time, the CHPD model based on MISOCP is solved using the 
proposed adaptive algorithm. 
To effectively enhance the relaxation quality while balancing computational efforts, 
the adaptive strategy sequentially divides the domains of variables 
belonging to the most severely violated bilinear constraints 
into smaller partitions for tighter piecewise polyhedral relaxation. 
At each iteration, an updated CHPD model in MISOCP is formulated and solved 
to find a relaxed solution. Then, a feasible solution is determined 
according to the mass flow rates of the relaxed solution.  
Note that the previous solution can be saved to  
provide a warm start for solving the MISOCP problem in the next iteration,  
expediting the pruning and bounding process.
The termination criteria can be tuned according to 
specific operator requirements, including a maximum count of iterations,
a time limit, or a desirable objective bound to stop.

\section{Case Studies}

\subsection{System Configuration}

\begin{figure}[htb]
	\centering
	\includegraphics[width=\columnwidth]{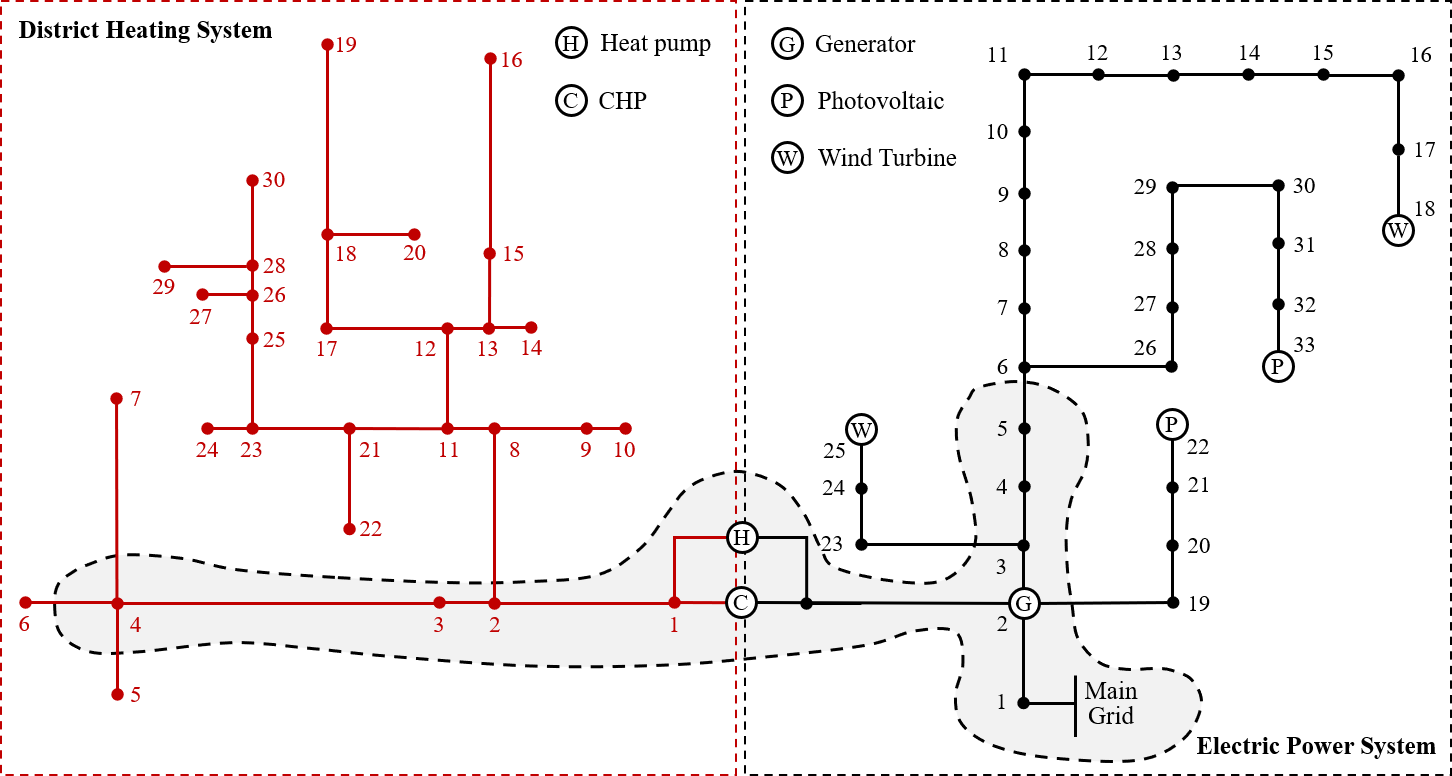}
	\caption{Diagram of the test integrated electricity and heat system.}
	\label{fig:system}
\end{figure}

The proposed convex CHPD model and adaptive solution algorithm are tested 
on an integrated electricity and heat system, 
composed of a 33-bus 12.66 kV EPS \cite{baran1989network} 
and a 30-node DHS \cite{palsson1999equivalent}.
Fig. \ref{fig:system} shows the test system with locations of 
the CHP, heat pump, gas-fired generator, wind turbines and photovoltaic arrays.
The DHS contains 29 pipes and 17 heat-exchangers with a total length of 6.6 km. 
The peak electricity and heat loads are 3.72 MW and 11.44 MW, respectively.

\subsection{Accuracy Validation of STD Model}

\begin{figure}[htb]
	\centering
	\subfloat[$m_{12}=150$ kg/s.]{
		\includegraphics[width=0.80\columnwidth]{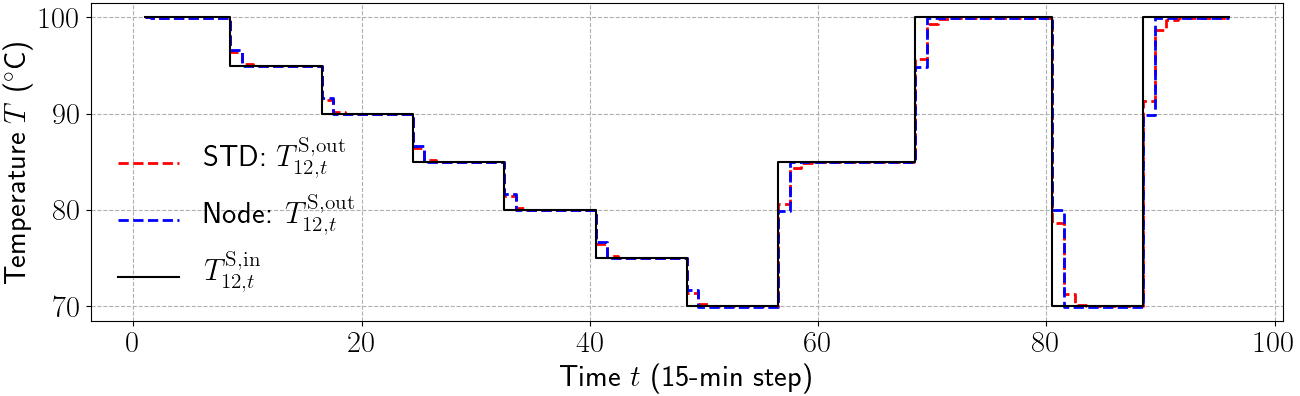}
		\label{fig:TempLarge}} \\
	\subfloat[$m_{12}=30$ kg/s.]{
		\includegraphics[width=0.80\columnwidth]{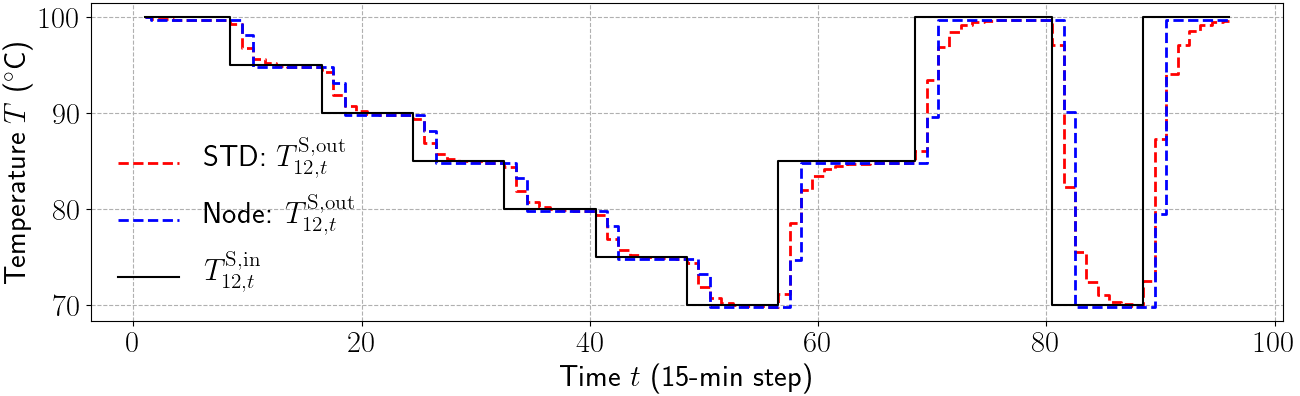}
		\label{fig:TempSmall}}
	\caption{Temperature dynamics in pipe $(1,2)$.}
	\label{fig:TempPipe}
\end{figure}

The accuracy of the proposed STD model is verified in 
comparison with the node method by simulating temperature dynamics in DHS 
with different mass flow rates. 
Fig. \ref{fig:TempPipe} shows the inlet temperature and the outlet temperature 
of pipe $(1,2)$
calculated by the node method (denoted as $T^{\mathrm{S,out}}_\mathrm{node}$)
and the STD model (denoted as $T^\mathrm{S,out}_\mathrm{STD}$).
The inlet temperature at node 1 is changed manually in a step-wise manner
to demonstrate corresponding temperature changes at the outlet.
Two scenarios ($m_{12}=150$ kg/s and $m_{12}=30$ kg/s) are given to 
evaluate the impacts of mass flow rates on time delays.
Overall, deviations between $T^\mathrm{S,out}_\mathrm{STD}$ 
and $T^{\mathrm{S,out}}_\mathrm{node}$ are small and STD  
provides good estimation of temperature dynamics.
In addition, time delays of heat delivery can also be captured 
by the proposed STD model, as illustrated in Fig. \ref{fig:TempSmall},
where a lower flow rate of 30 kg/s slows the heat delivery in this pipeline. 

\begin{figure}[htb]
	\centering
	\subfloat[Node 2.]{
		\includegraphics[width=0.80\columnwidth]{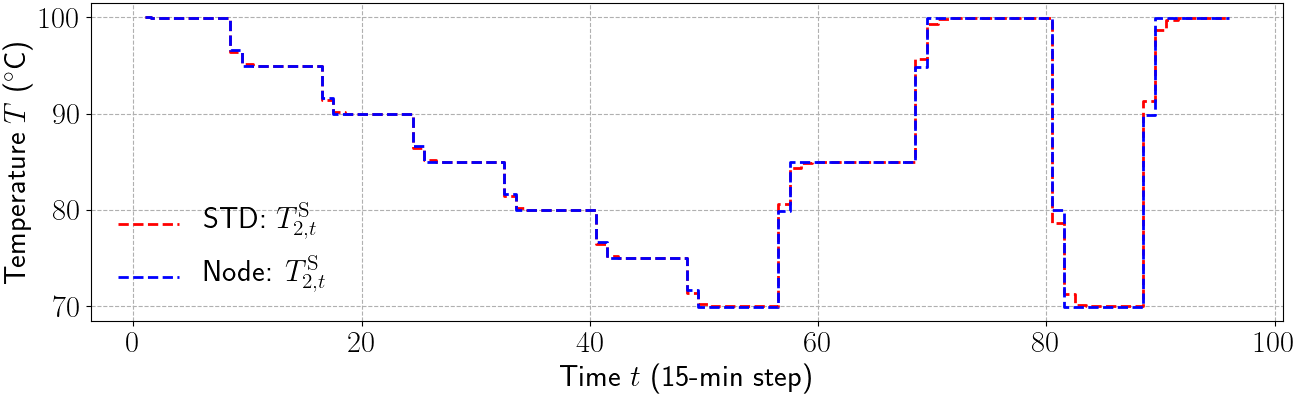}
		\label{fig:STDvsNode1}} \\
	\subfloat[Node 6.]{
		\includegraphics[width=0.80\columnwidth]{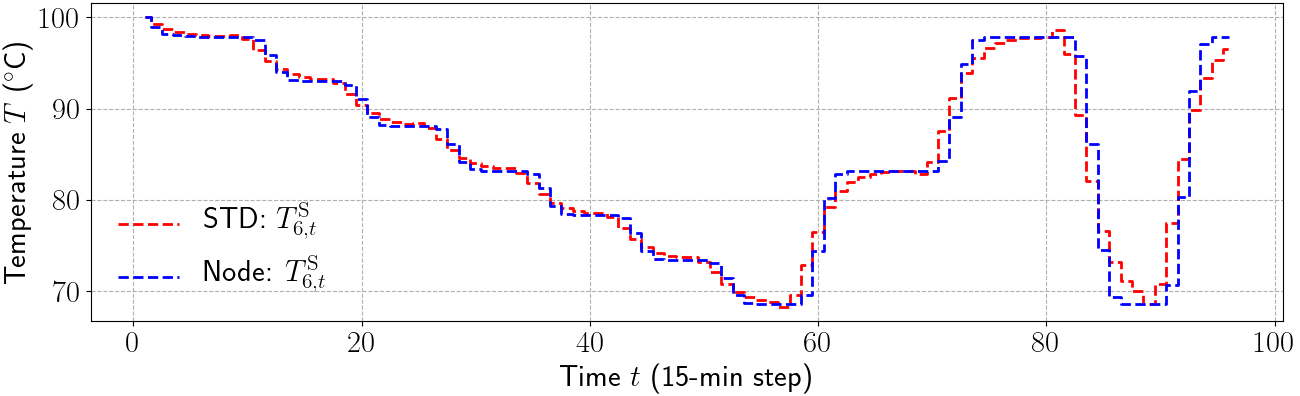}
		\label{fig:STDvsNode2}} \\
	\subfloat[Node 30.]{
		\includegraphics[width=0.80\columnwidth]{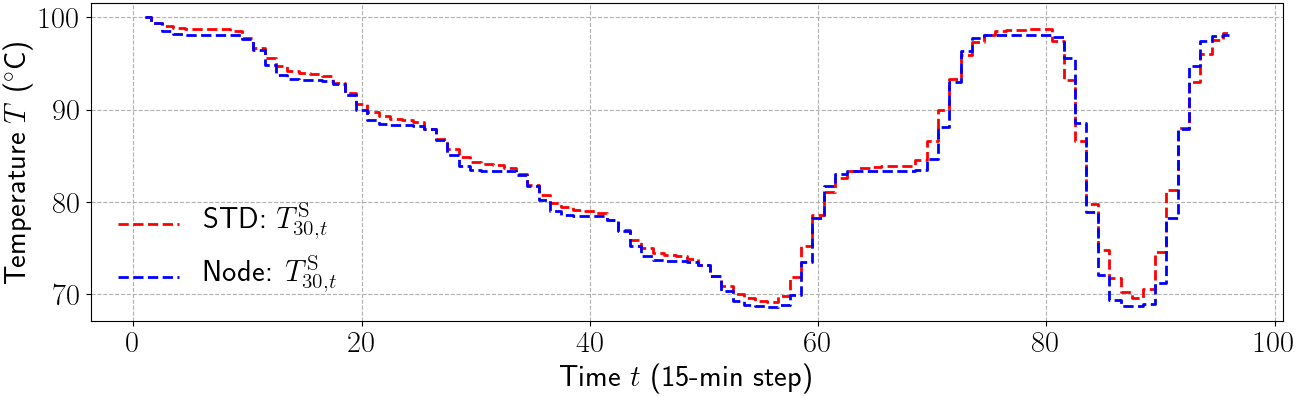}
		\label{fig:STDvsNode3}} \\
	\caption{Temperature changes in DHS based on the STD and the node method.}
	\label{fig:TempSTD}
\end{figure}

Fig. \ref{fig:TempSTD} illustrates the temperature changes in the whole DHS 
determined by the proposed STD model and the original node method.
Nodes 2, 6 and 30 are selected for demonstration.
The STD model tends to smooth sudden step changes of temperature,
which can be observed by comparing 
temperature profile from $t=0$ to $t=60$ and that at $t=80\sim96$. 
The relative error of temperature estimation is 0.68\% on average.
In summary, Fig. \ref{fig:TempSTD} illustrates that 
the STD model can capture temperature dynamics in the whole network 
with desirable accuracy.

\subsection{Comparison Between Convex and Constant-Flow CHPD}

This section demonstrates the performance of the proposed convex CHPD model
and CHPD based on MISOCP. 
The constant-flow CHPD approach is the benchmark, where hydraulic conditions 
are given by pre-defined mass flow rates. 
In addition to the full system, we include a prototype test based on 
a hypothetical small system.
The test system is shown in gray-shaded area in Fig. \ref{fig:system},
consisting of a 5-bus EPS and a 4-node DHS. 
A photovoltaic array and a wind turbine are installed at bus 4 and bus 5, respectively.
Peak electricity and heat demands are 1.55 MW and 2.46 MW, respectively.
Capacities of heat and power generation units are reduced proportionally.
The following simulation results are obtained based on a PC 
with Intel Core i7-8750H @2.2GHz, 16GB RAM. 
Algorithms are implemented using Python 3.7.5 
in conjunction with Gurobi 9.0.0 \cite{gurobi}.

Table \ref{tab:ConvexConstant} shows final costs and computing time derived from
different methods.
Here, $n$ indicates the number of partitions for all variable domains 
associated with bilinear constraints. 
Hence $n=1$ refers to convex CHPD models without piecewise polyhedral relaxation.
CHPD-MISOCP with $n\geq2$ indicates dispatch models based on
globally uniform variable partitioning without implementation of the adaptive algorithm.
In Table \ref{tab:ConvexConstant}, the convex CHPD method outperforms 
the constant-flow CHPD by at least 3.56\% in the large system 
and 1.04\% in the small system with respect to the total cost.
Cost reduction, i.e., solution optimality, can be further improved 
by enabling piecewise convex relaxation.
A cost saving of \$369 (3.86\%) can be achieved with $n=2$ for the large test system
and \$116 (4.46\%) for the small test system compared to $n=1$.

The computational efficiency of the convex CHPD model is excellent and approximately
identical to the constant-flow CHPD approach, since both are SOCP problems.
A convex CHPD problem for the large test system can be solved within 1 second 
with 2472 second-order cone constraints.
In general, the CHPD model based on MISOCP is much more time-consuming 
than convex CHPD schemes as expected
due to the incorporation of binary variables and
extensive space search for branching and bounding. 
For instance, the CHPD-MISOCP model of the large test system  
has 10752 binary variables with $n=2$
and 16128 binaries with $n=3$. 
For the test system presented in this paper,
qualified solutions for CHPD-MISOCP are not available within 60 minutes
if $n$ is greater than 2.
Here, the optimality gap of the mixed-integer programming solver is set to 0.01\%.
\begin{table}[htb]
	\footnotesize
	\centering
	\caption{Comparison Between Convex and Constant-Flow CHPD Models}
	\label{tab:ConvexConstant}
	\begin{threeparttable}
	\begin{tabular}
		{@{\hspace{6pt}}c@{\hspace{6pt}}c@{\hspace{6pt}}c
			@{\hspace{6pt}}c@{\hspace{6pt}}c@{\hspace{6pt}}}
	\toprule 
	System & Method & Setting & Cost (\$) & Time (s) \\ \midrule
	\multirow{6}{*}{\shortstack{5-bus EPS\\+\\4-node DHS}} 
	& Convex CHPD & $n=1$ & 2580.89 & 0.19 \\ 
	& CHPD-MISOCP & $n=2$ & 2491.89 & 141.53 \\ 
	& CHPD-MISOCP & $n=3$ & /\tnote{1} & $>$3600.00 \\ \cline{2-5}
	& \multirow{3}{*}{\shortstack{Constant-flow\\CHPD}} 
		& $m=100$ kg/s & 2625.73 & 0.03 \\
	&  & $m=80$ kg/s & 2608.13 & 0.02 \\
	&  & $m=60$ kg/s & -\tnote{2} & - \\ \midrule
	\multirow{6}{*}{\shortstack{33-bus EPS\\+\\30-node DHS}} 
	& Convex CHPD & $n=1$ & 9230.79 & 0.76 \\ 
	& CHPD-MISOCP & $n=2$ & 9202.23 & 2165.01 \\ 
	& CHPD-MISOCP & $n=3$ & / & $>$3600.00 \\ \cline{2-5}
	& \multirow{3}{*}{\shortstack{Constant-flow\\CHPD}} 
		& $m=200$ kg/s & 9746.15 & 0.11 \\
	&  & $m=175$ kg/s & 9571.48 & 0.10 \\
	&  & $m=150$ kg/s & - & - \\ \bottomrule
	\end{tabular}
	\begin{tablenotes}
		\item[1] / means no solution found within 1 hour.
		\item[2] - means no feasible solution found.
	\end{tablenotes}
	\end{threeparttable}  
\end{table}

\begin{figure}[htb]
	\centering
	\subfloat[Power generation.]{
		\includegraphics[width=0.47\columnwidth]{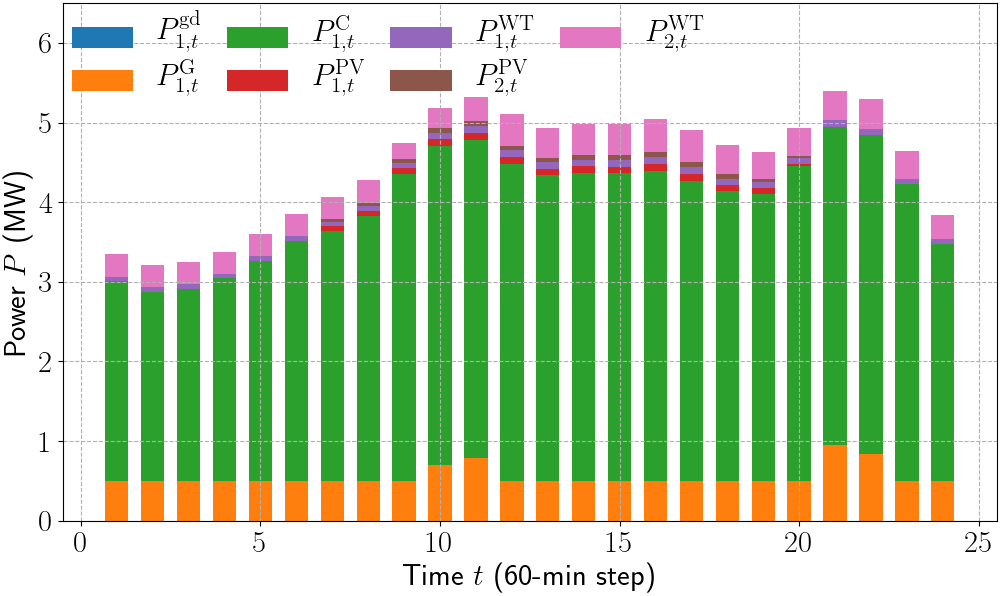}
		\label{fig:PowGenCons}}
	\subfloat[Power loads.]{
		\includegraphics[width=0.47\columnwidth]{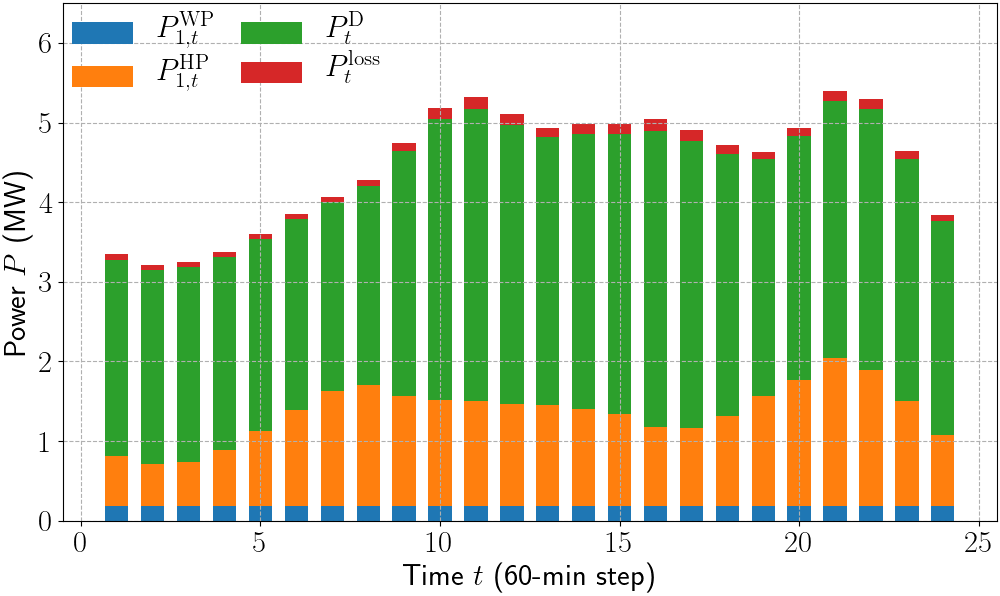}
		\label{fig:PowLoadCons}} \\
	\subfloat[Heat balance.]{
		\includegraphics[width=0.47\columnwidth]{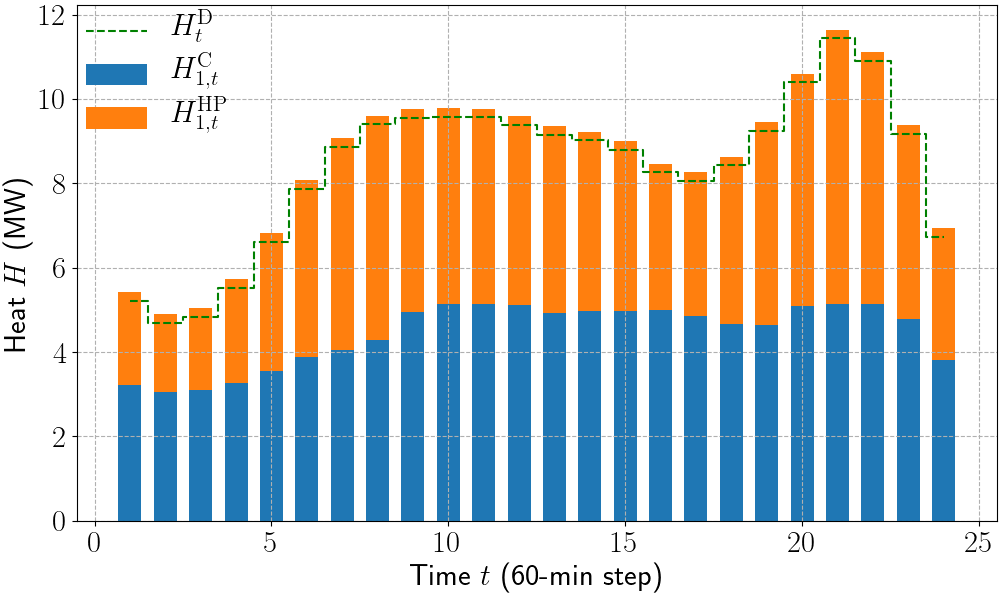}
		\label{fig:HeatBalCons}}
	\subfloat[Mass flow rates.]{
		\includegraphics[width=0.47\columnwidth]{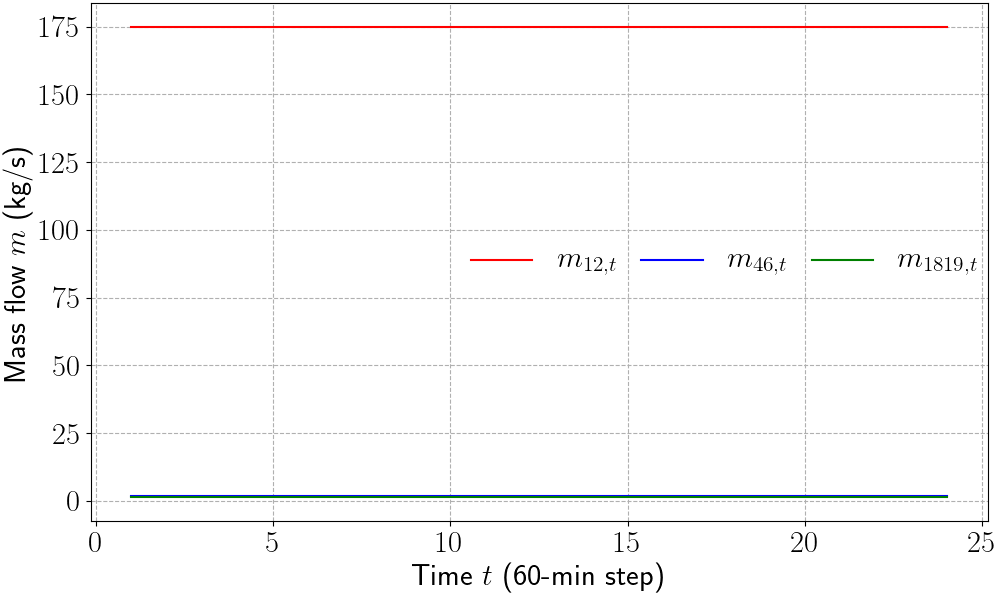}
		\label{fig:MassFlowCons}}
	\caption{Scheduling strategies of the constant-flow CHPD model}
	\label{fig:ScheCons}
\end{figure}

\begin{figure}[htb]
	\centering
	\subfloat[Power generation.]{
		\includegraphics[width=0.47\columnwidth]{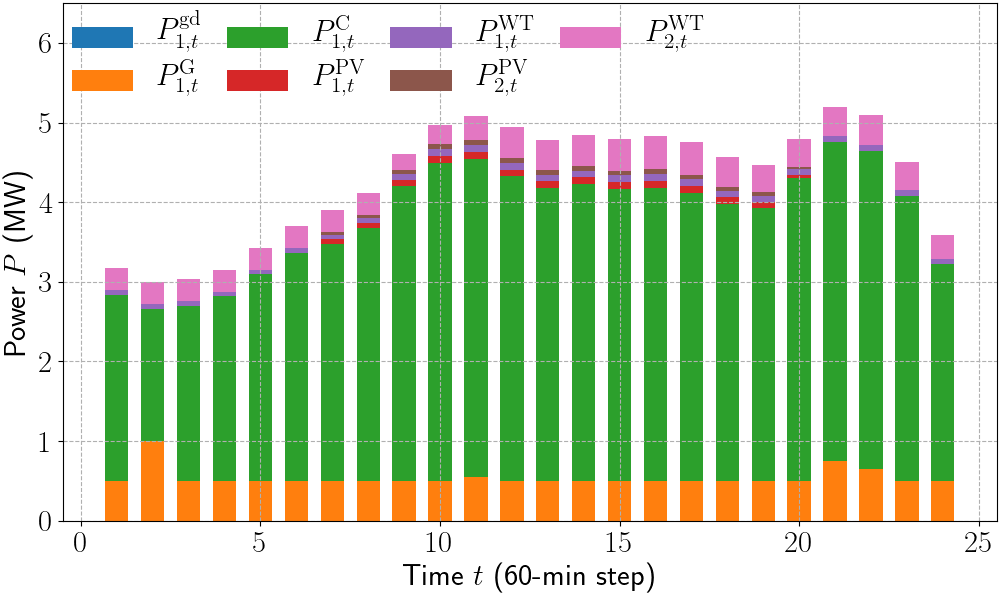}
		\label{fig:PowGenConvex}}
	\subfloat[Power loads.]{
		\includegraphics[width=0.47\columnwidth]{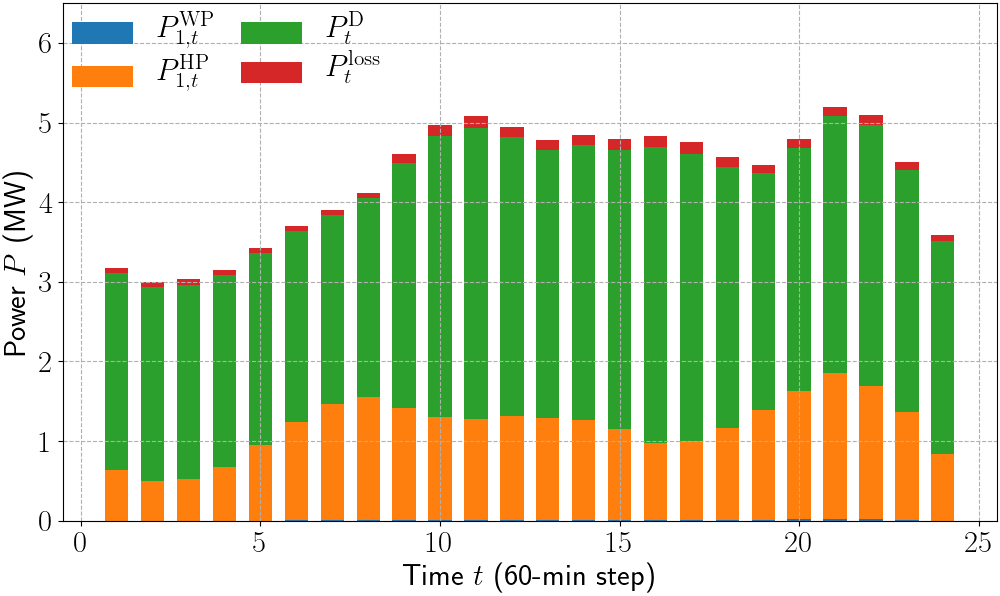}
		\label{fig:PowLoadConvex}} \\
	\subfloat[Heat balance.]{
		\includegraphics[width=0.47\columnwidth]{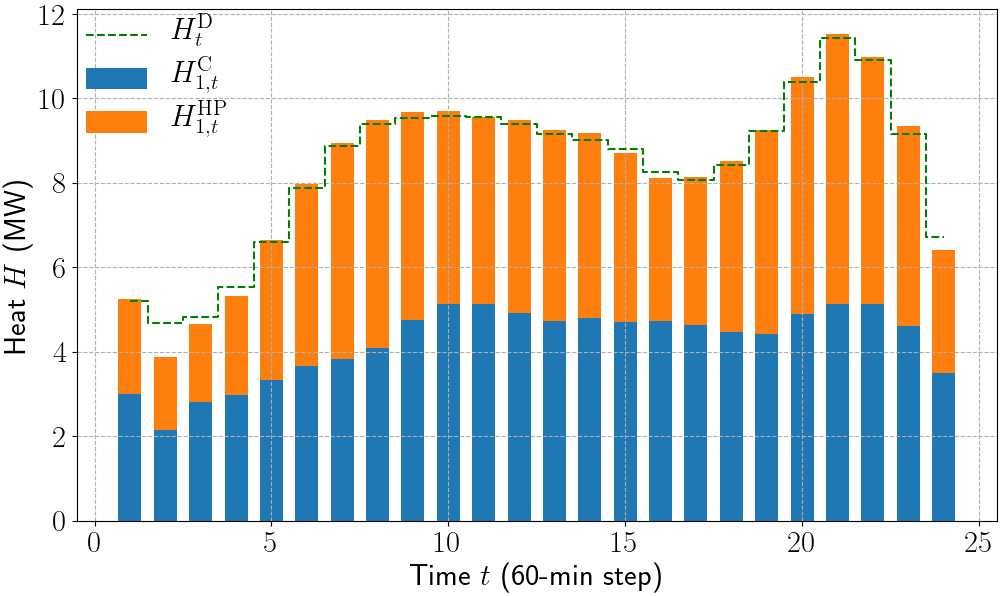}
		\label{fig:HeatBalConvex}}
	\subfloat[Mass flow rates.]{
		\includegraphics[width=0.47\columnwidth]{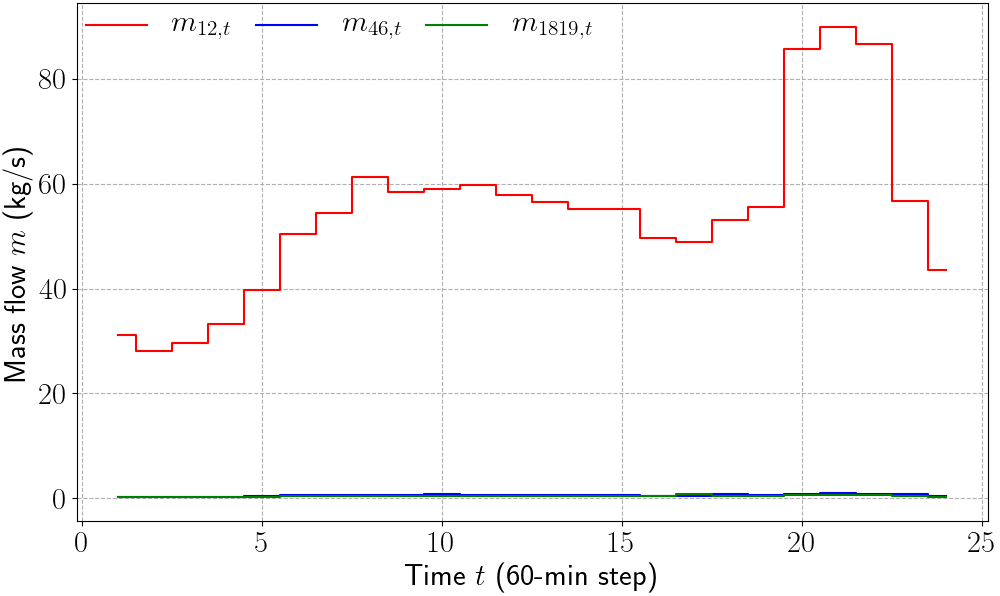}
		\label{fig:MassFlowConvex}}
	\caption{Scheduling strategies of the CHPD-MISOCP model}
	\label{fig:ScheConvex}
\end{figure}

Scheduling strategies from the CHPD model with $m=175$ kg/s
and the relaxed CHPD-MISOCP model with $n=2$ are given in 
Fig. \ref{fig:ScheCons} and Fig. \ref{fig:ScheConvex}, respectively.
The CHP supplies a large proportion of electricity and heat loads due to 
the low-cost nature of co-generation. 
Remaining heat demands are satisfied by the heat pump.
Unlike the electricity loads that need to be instantaneously balanced,
there is a mismatch between heat generation and heat demands 
in Fig. \ref{fig:HeatBalCons} and Fig. \ref{fig:HeatBalConvex}
owing to the storage capability of networked heating pipelines.
Renewable power of wind turbines and photovoltaic panels are integrated into 
the system to reduce generation cost of electricity.
One of the main differences between constant-flow scheduling strategies and 
those from the CHPD-MISOCP approach is that
the mass flow rates are regulated to a lower level to reduce electricity 
consumption of water pumps.
Fig. \ref{fig:MassFlowCons} and \ref{fig:MassFlowConvex} show mass flow rates
derived by the constant-flow model and the proposed CHPD-MISOCP model.
As a result, the electricity consumption for water pumps
(about 4.5 MWh) can be saved without violating any constraints or causing load shedding,
which demonstrates the effectiveness of CHPD 
in achieving joint cost savings for integrated electricity and heat systems.
In summary, the proposed convex-relaxation-based CHPD model  
produces high-quality operational strategies with significant
cost reductions.

\subsection{Effectiveness of Adaptive Solution Algorithm}

Table \ref{tab:Adaptive} demonstrates the effectiveness of the proposed 
adaptive solution algorithm compared to the NLP solver in Gurobi 9.0.0.
The termination conditions for both algorithms are time limits,
namely, 10-, 30, and 60-minutes, respectively.
Gurobi exploits a search scheme based on spatial branch-and-bound (sBB), 
where branching occurs on one variable at each iteration to 
generate sub-problems for further bounding and pruning.
In Table \ref{tab:Adaptive}, the final cost of optimized strategies obtained 
from the proposed algorithm is lower than that of sBB by 
1.72\% in the large test system and 1.04-2.95\% in the small system
based on the same time limit.
Besides, by comparing Table \ref{tab:ConvexConstant} and \ref{tab:Adaptive},
we see that the adaptive solution algorithm produces a better solution 
than CHPD-MISOCP based on uniform partitioning.
For instance, the adaptive algorithm generates a solution of \$9017.05 in 30 minutes, 
while CHPD-MISOCP takes 2165.01 seconds to derive a solution that is 2.05\% more 
expensive (\$9202.23). 
The improvement on the small test system can be as much as 3.85\% by switching to the dynamic 
bivariate partitioning.
\begin{table}[htb]
	\footnotesize
	\centering
	\caption{Comparison Between Adaptive Algorithm and NLP Solver}
	\label{tab:Adaptive}
	\begin{threeparttable}
	\begin{tabular}
		{@{\hspace{6pt}}c@{\hspace{6pt}}c
			@{\hspace{6pt}}c@{\hspace{6pt}}c@{\hspace{6pt}}c@{\hspace{6pt}}}
	\toprule 
	\multirow{2}{*}{System} & \multirow{2}{*}{Method} & Cost (\$) & Cost (\$) & Cost (\$) \\
	& & $\bar{t}=$600s & $\bar{t}=$1800s & $\bar{t}=$3600s \\ \midrule
	\multirow{2}{*}{\shortstack{5-bus EPS+\\4-node DHS}} 
	& Adaptive & 2439.73 & 2439.73 & 2392.59 \\ 
	& sBB & 2505.58 & 2465.40 & 2465.40 \\ \midrule 
	\multirow{2}{*}{\shortstack{33-bus EPS+\\30-node DHS}} 
	& Adaptive & 9191.15 & 9017.05 & 9017.05 \\ 
	& sBB & -\tnote{1} & 9174.67 & 9174.67 \\ \bottomrule
	\end{tabular}
	\begin{tablenotes}
		\footnotesize
		\item[1] - means no feasible solution found.
	\end{tablenotes}
	\end{threeparttable} 
\end{table}

\begin{figure}[htb]
	\centering
	\includegraphics[width=0.70\columnwidth]{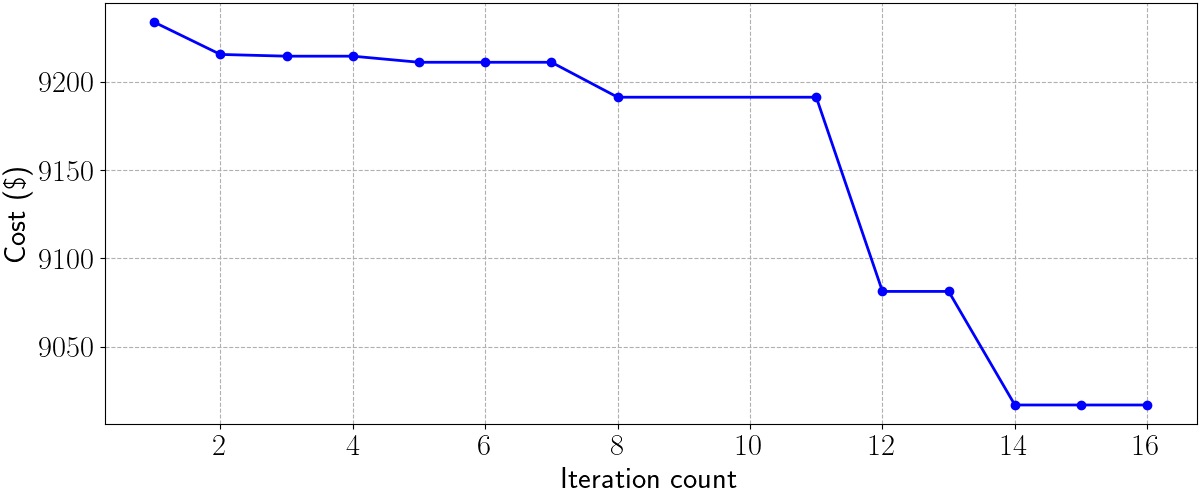}
	\caption{Evolution of objective function values in iterations.}
	\label{fig:Convergence}
\end{figure}

Fig. \ref{fig:Convergence} shows the changes of operational costs by iterations
using the proposed adaptive algorithm with a time limit of 30 minutes. 
Note that the total cost obtained at each iteration is the best solution, i.e.,
the lower bound of existing solutions in previous iterations.
Hence, the proposed adaptive algorithm can improve the solution quality 
successively by increasing the number of partitions. 

\section{Conclusion}
This paper proposes a novel convex CHPD model based on model simplification and 
constraint relaxation. 
The proposed model is universally applicable without any assumptions 
on operating regimes of DHS, and avoids non-deterministic 
and extensive parameter tuning required in heuristic-based algorithms.
A simplified thermal dynamic model is developed to alleviate logic complexity 
in the node method. Furthermore, an adaptive solution algorithm is proposed 
based on dynamic bivariate partitioning to sequentially tighten 
piecewise convex relaxation for improvements in solution quality.
Case study results based on a 33-bus EPS and a 30-node DHS 
illustrate that the proposed convex CHPD method can derive high-quality solutions
with significant cost reductions at excellent computational efficiency.
Moreover, the adaptive solution algorithm 
provide an effective way to further improve the solution quality 
with desirable computational efficiency.  



\ifCLASSOPTIONcaptionsoff
  \newpage
\fi



\bibliographystyle{IEEEtran}
\bibliography{ConvexCHPD}

\end{document}